\documentclass[journal]{IEEEtran}
\ifCLASSOPTIONcompsoc
  \usepackage[nocompress]{cite}
\else
  \usepackage{cite}
\fi
\ifCLASSINFOpdf
\else
\fi
\usepackage{cite}
\usepackage{amsmath,amssymb,amsfonts}
\usepackage{graphicx}
\usepackage{textcomp}
\usepackage{xcolor}
\usepackage{comment}
\usepackage{enumitem}
\usepackage{multirow}
\usepackage{booktabs}
\usepackage{soul}
\usepackage{amsmath,bm}
\usepackage{xspace}
\usepackage{pifont}
\usepackage{lipsum}
\usepackage{graphics}
\usepackage{algpseudocode}
\usepackage{amsmath}
\usepackage{graphicx}
\usepackage{subfigure}
\usepackage{pifont}
\usepackage{mathtools}
\usepackage{hyperref}
\usepackage{ragged2e}
\usepackage{algorithmicx,algorithm}
\usepackage{cite}
\usepackage{amsmath,amssymb,amsfonts}
\usepackage{graphicx}
\usepackage{textcomp}
\usepackage{xcolor}
\usepackage{CJK}
\usepackage{indentfirst}
\usepackage{flushend}
\renewcommand{\raggedright}{\leftskip=0pt \rightskip=0pt plus 0cm}

\renewcommand{\raggedright}{\leftskip=0pt \rightskip=0pt plus 0cm} 
\def\BibTeX{{\rm B\kern-.05em{\sc i\kern-.025em b}\kern-.08em
		T\kern-.1667em\lower.7ex\hbox{E}\kern-.125emX}}

\begin{document}
	
\title{HyCA: A Hybrid Computing Architecture for Fault Tolerant Deep Learning}

\author{
Cheng~Liu, ~\IEEEmembership{Member,~IEEE,} 
Cheng~Chu,
Dawen~Xu,
Ying~Wang, ~\IEEEmembership{Member,~IEEE,} 
Qianlong~Wang, \\
Huawei Li, ~\IEEEmembership{Senior~Member,~IEEE,} 
Xiaowei Li, ~\IEEEmembership{Senior~Member,~IEEE,} 
Kwang-Ting Cheng, ~\IEEEmembership{Fellow,~IEEE}

\IEEEcompsocitemizethanks{
{\IEEEcompsocthanksitem Manuscript received May 24, 2021; revised August 18, 2021; accepted October 13, 2021. Date of publication
XXXX XX, 202X; This article was presented in part
at The 38th IEEE International Conference on Computer Design (ICCD), October 18–21, 2020.\\}
{\IEEEcompsocthanksitem Cheng Liu, Ying Wang, and Xiaowei Li are with SKLCA, Institute of Computing Technology, Chinese Academy of Sciences, Beijing 100080, China. E-mail: liucheng@ict.ac.cn\\}
{\IEEEcompsocthanksitem Huawei Li is with both SKLCA, Institute of Computing Technology, Chinese Academy of Sciences, Beijing 100080, China and Peng Cheng Laboratory, Shenzhen 518055, China.\\}
{\IEEEcompsocthanksitem Dawen Xu, Qianlong Wang and Cheng Chu are with both Hefei University of Technology, Hefei 230009, China and SKLCA, Institute of Computing Technology, Chinese Academy of Sciences, Beijing 100080, China. Email: chucheng@mail.hfut.edu.cn\\}
{\IEEEcompsocthanksitem Kwang-Ting Cheng is with with Department of Computer Science and Engineering, The Hong Kong University of Science and Technology, Hong Kong 999077, China.\protect\\}
}


}

\IEEEtitleabstractindextext{%
	
\begin{abstract}
\raggedright Hardware faults on the regular 2-D computing array of a typical deep learning accelerator (DLA) can lead to dramatic prediction accuracy loss. Prior redundancy design approaches typically have each homogeneous redundant processing element (PE) to mitigate faulty PEs for a limited region of the 2-D computing array rather than the entire computing array to avoid the excessive hardware overhead. However, they fail to recover the computing array when the number of faulty PEs in any region exceeds the number of redundant PEs in the same region. The mismatch problem deteriorates when the fault injection rate rises and the faults are unevenly distributed. To address the problem, we propose a hybrid computing architecture (HyCA) for fault-tolerant DLAs. It has a set of dot-production processing units (DPPUs) to recompute all the operations that are mapped to the faulty PEs despite the faulty PE locations. According to our experiments, HyCA shows significantly higher reliability, scalability, and performance with less chip area penalty when compared to the conventional redundancy approaches. Moreover, by taking advantage of the flexible recomputing, HyCA can also be utilized to scan the entire 2-D computing array and detect the faulty PEs effectively at runtime. 
\end{abstract}

}

\maketitle
\IEEEdisplaynontitleabstractindextext
\IEEEpeerreviewmaketitle

\section{Introduction} \label{sec:intro}
The great success of deep learning motivates the deployment of deep learning in numerous domains of applications. Many of the applications such as autonomous driving and drones \cite{fink2019deep, tzelepi2017human}, and intelligent medical monitoring and treatment \cite{esteva2019guide} are closely related to the safety of human beings and are mission-critical. When deep learning models are applied in these applications, the reliability of the execution is of vital importance and must be considered comprehensively \cite{banerjee2019towards, jha2019ml}. Otherwise, the unexpected inference predictions may lead to catastrophic consequences \cite{jenihhin2019challenges}. While the deep learning models are increasingly implemented on customized deep learning accelerators (DLAs) for the sake of both higher performance and energy efficiency \cite{chen2014dadiannao}, the reliability of the model execution dramatically depends on the underlying accelerators \cite{xu2020hybrid, reagen2016minerva}. At the same time, DLAs fabricated with continuously shrinking semiconductor technologies are more likely to suffer manufacture defects and become more sensitive to the working conditions such as the large temperature variation than before \cite{impact2011dixit}, which may cause hardware faults and incur considerable prediction accuracy loss. Thereby, resilient DLAs are indispensable for reliable inference and are highly demanded by the mission-critical AI applications \cite{mittal2020survey}.

DLAs typically consist of a large regular computing array which can either be a systolic array or a plain mesh array\cite{jouppi2017datacenter, Chen2016Eyeriss}, and a set of on-chip buffers used for input features, output features, and weights. While the on-chip buffers implemented with SRAMs can be usually protected effectively with error correction code (ECC), we mainly focus on the reliability design of the regular computing array in this work. As each processing element (PE) in the computing array can be used for the calculation of multiple features in different network layers, faults in a single PE may cause multiple faulty computing results during the deep learning model execution. Thereby, they may result in considerable accuracy degradation according to our experiments in Section \ref{sec:motivation}. 

To mitigate the hardware faults in the 2-D computing array of DLAs, researchers have proposed a number of fault-tolerant design approaches from distinct angles. These approaches can be roughly divided into two categories. The first category mainly exploits the inherent fault-tolerance of the neural network models by retraining the neural network models specifically for the faulty computing array without any modification or with minor modification to the computing array \cite{deng2015retraining, zhang2019fault, xu2019resilient, li2019squeezing}. Although these approaches induce negligible hardware overhead and can even be applied to many off-the-shelf accelerators, the model retraining is required for each specific fault configuration, and the retraining, especially for large data sets and deep learning models, can be rather expensive. For instance, a critical neural network model applied in automotive systems must go through a series of standard certification tests before the modification can be accepted \cite{validation2019Ebert}. The cost of the certification test is nontrivial and time-consuming. Moreover, the prediction accuracy of the retrained models depends on both the model structures and the specific fault configurations. There is no guarantee that the model retraining can always maintain the original model accuracy and fulfill the requirements of mission-critical AI applications for all the different fault configurations by design.

The second category aims to recover the faulty computing array with redundant PEs. While the conventional approaches such as dual modular redundancy (DMR) and triple modular redundancy (TMR) require substantial hardware resources, researchers proposed a variety of relaxed redundancy approaches. The basic idea is to have each redundant PE to recover any faulty PE in a limited region of the computing array while the region can be a row, a column, or both row and column \cite{takanami2012built} \cite{takanami2017built}, which essentially limits the sharing of the redundant PEs and reduces the hardware resource consumption significantly compared to the DMR and TMR approaches. When the faulty PEs are evenly distributed across the computing array, the faulty PEs can be probably mitigated. Nevertheless, the faults may not be evenly distributed across the computing array and these approaches fail to recover the computing array even when the number of redundant PEs exceeds the number of the faulty PEs in the computing array. In this case, the DLAs will not be fully functional or degrade dramatically if the faulty PEs are discarded. In summary, there is still a lack of resilient computing array architectures for DLAs that allow unmodified deep learning model execution and can tolerate various fault distribution at the same time.  

To address the problems, we propose a hybrid computing architecture (HyCA) for the fault-tolerant DLAs. Unlike prior works in which each homogeneous redundant PE can only protect a small subset of PEs in the 2-D computing array, we have a set of dot-production processing units (DPPUs) implemented as redundancy targeting at any faulty PEs in the 2D computing array. DPPUs can have all the output feature calculation that is mapped to the faulty PEs in the 2-D computing array recomputed. The locations of the faulty PEs in the 2-D computing array will not affect the recomputing. When the DPPU computing power is lager than the recomputing requirements, the neural network execution can be recovered without any performance penalty nor accuracy loss. When the DPPU computing power is insufficient, we may either stall the 2-D computing array or discard some of the faulty PEs, which will degrade the performance of the DLAs. In addition, the DPPUs can also be utilized to sequentially scan and check the computing results of all the PEs in the 2-D computing array such that faulty PEs in the 2-D computing array can also be detected at runtime. 

The contributions of this work are summarized as follows. The second and the third contributions are mainly extended on top of DPPU proposed in the conference paper \cite{xu2020hybrid}.
\begin{enumerate}
    \item We propose a hybrid computing architecture for fault-tolerant DLAs. It has a set of DPPUs seated along with the regular 2-D computing array to recompute all the operations mapped to the faulty PEs in arbitrary locations of the 2-D computing array. DPPUs work concurrently with the 2-D computing array and impose neither the accuracy penalty nor the performance penalty to the deployed neural network models when the number of faulty PEs is smaller than the DPPU size. When the number of faulty PEs exceeds the DPPU size, DPPUs can mitigate the faulty PEs that are more critical to the performance to achieve graceful performance degradation.
    
    \item Meanwhile, we propose a grouped DPPU and banked register files with shifting support to decouple the 2-D computing array size and the DPPU size such that proposed HyCA can be scaled for distinct design constraints.
    
    \item By taking advantage of the DPPU recomputing capability, we propose to have part of the grouped DPPU to check the computing results of PEs in the 2-D computing array. Moreover, instead of checking the final computing results, we propose to check only the partial computing results and scan each PE sequentially such that the faults in the 2-D computing array can be detected efficiently at runtime.
    
    \item According to our experiments, HyCA achieves both higher performance and reliability under various fault injection setups with less chip area penalty when compared to the classical redundancy design approaches. In addition, HyCA with the grouped DPPU structures and banked register files exhibits scalable performance and prompt fault detection capability.
    
\end{enumerate}


\section{Related Work} \label{sec:relatedwork}
With the increasing adoption of deep learning in mission-critical applications, such as autonomous driving and drones, the reliability of DLAs widely utilized for the deep learning processing becomes critical and attracts a lot of attentions of the researchers recently \cite{mittal2020survey} \cite{xu2021reliability} \cite{xu2020persistent} \cite{ning2020ftt}. To analyze the influence of hardware faults on the deep neural network models, the authors in \cite{zhang2019fault} conducted comprehensive experiments and the experiment results show that hardware faults can lead to considerable prediction accuracy drop. For the TIMIT speech recognition task, the accuracy drops from 74.13$\%$ to 39.69$\%$. The accuracy loss is relevant to various design factors including the quantization, data format, and network architecture. It remains rather challenging to ensure resilient deep neural network execution on DLAs.

To alleviate the influence of hardware faults on neural network predictions, many prior works \cite{error2018date} \cite{energy2018kim} \cite{axtrain2018he} took advantage of the intrinsic fault-tolerance of neural network models by retraining the models for a specific fault configuration such that hardware faults can be compensated by the retrained models. Xu et al. \cite{xu2019resilient} proposed an on-accelerator retraining framework to obtain models that can tolerate the random hardware faults. Li et al. \cite{li2019squeezing} and Wang et al. \cite{wang2017resilience} proposed to employ the model retraining for DLAs with overclocking which may incur timing errors. To train resilient deep learning models, He et al. \cite{axtrain2018he} revised the loss function to obtain models that are less sensitive to the hardware faults. Unlike the above methods, Zhang et al. \cite{analyzing2018vts} \cite{zhang2019fault} proposed to bypass the faulty PEs in the computing array with zeros or other constant values such that the faults are more easier to tolerate via retraining. Although the retraining works for many fault configurations, there is still no guarantee that the retrained models can maintain the prediction accuracy for the target mission-critical applications because of the huge number of different fault configurations. In addition, the retraining can be rather expensive especially for large datasets and models, and is required for each specific fault configuration, which further limits the adoption of the retraining approaches. Unlike the retraining-based approaches, Hanif et al. \cite{abdullah2020salvagednn} proposed a training-free mapping approach to alleviate the influence of permanent PE faults in the DLA. It leverages the saliency of the neural network parameters and opts to map the salient weights to the faulty PEs as much as possible such that the negative influence of the PE faults on the neural network models is reduced. However, it works only when the fault error rate is low and can be sensitive to the fault distribution.  

To enable unmodified deep learning model execution without accuracy loss, an intuitive approach is to develop reliable DLA architectures with the conventional TMR and DMR approaches to tolerate the hardware faults, but these approaches incur substantial hardware resource consumption. In this case, the authors in \cite{takanami2017built}\cite{takanami2012built}\cite{horita2000fault} proposed to add redundant PEs to the large regular homogeneous computing array and each redundant PE can be shared by a group of PEs with distinct redundancy sharing methods such as row redundancy (RR), column redundancy (CR), and diagonal redundancy (DR). For the RR and CR, each row/column of the PEs share the same set of redundant PEs. For the DR, both the row and column of PEs corresponding to a diagonal location in the computing array share the same set of redundant PEs. Since the redundant PEs are shared by a group of homogeneous PEs, hardware resource consumption can be greatly reduced compared to DMR and TMR. Nevertheless, the faulty PEs may not be evenly distributed or even clustered across the computing array \cite{stapper1983integrated}. As a result, these approaches may fail to recover the computing array when the number of faulty PEs in each shared region such as a row or a column of PEs exceeds the the number of shared redundant PEs in the region. Thereby, the utilization of the redundant PEs can be affected by the fault distribution dramatically. More redundant PEs must be designed to ensure reliable execution. Otherwise, the performance can degrade dramatically when the faulty PEs are discarded due to the insufficient redundant PEs. Thereby, more efficient computing array architectures are required for the highly resilient DLA designs. 

On top of the redundancy-based fault-tolerant DLA designs, there are also many other different fault-tolerant architecture designs. The authors in \cite{ozen2019sanity} \cite{zhao2020algorithm} proposed a spatial and temporal checksum to protect full connection and convolution layers in deep neural network models. Zhang et al. \cite{zhang2020sorting} proposed a parallel stochastic computing(SC)-based neural network accelerator purely using bitstream computation by fully exploiting the superior fault tolerance of SC mainly for ternary neural networks. Li et al. \cite{li2020soft} proposed an error detecting scheme to locate incorrect PEs of the DLA and gave an error masking method to achieve fault-tolerance. Hamid Reza Mahdiani etc. \cite{mahdiani2012relaxed} proposed to relax the fault-tolerance of the VLSI implementation by employing TMR to only the computation of the most important bits such that the hardware overhead is reduced and the critical path latency is improved. Nevertheless, these approaches either require model retraining or can be sensitive to the fault distribution. 
\section{Background and Motivation} \label{sec:motivation}
\begin{figure}
\setlength{\abovecaptionskip}{-1pt}
\setlength{\belowcaptionskip}{0pt}
	\center{\includegraphics[width=0.9\linewidth]{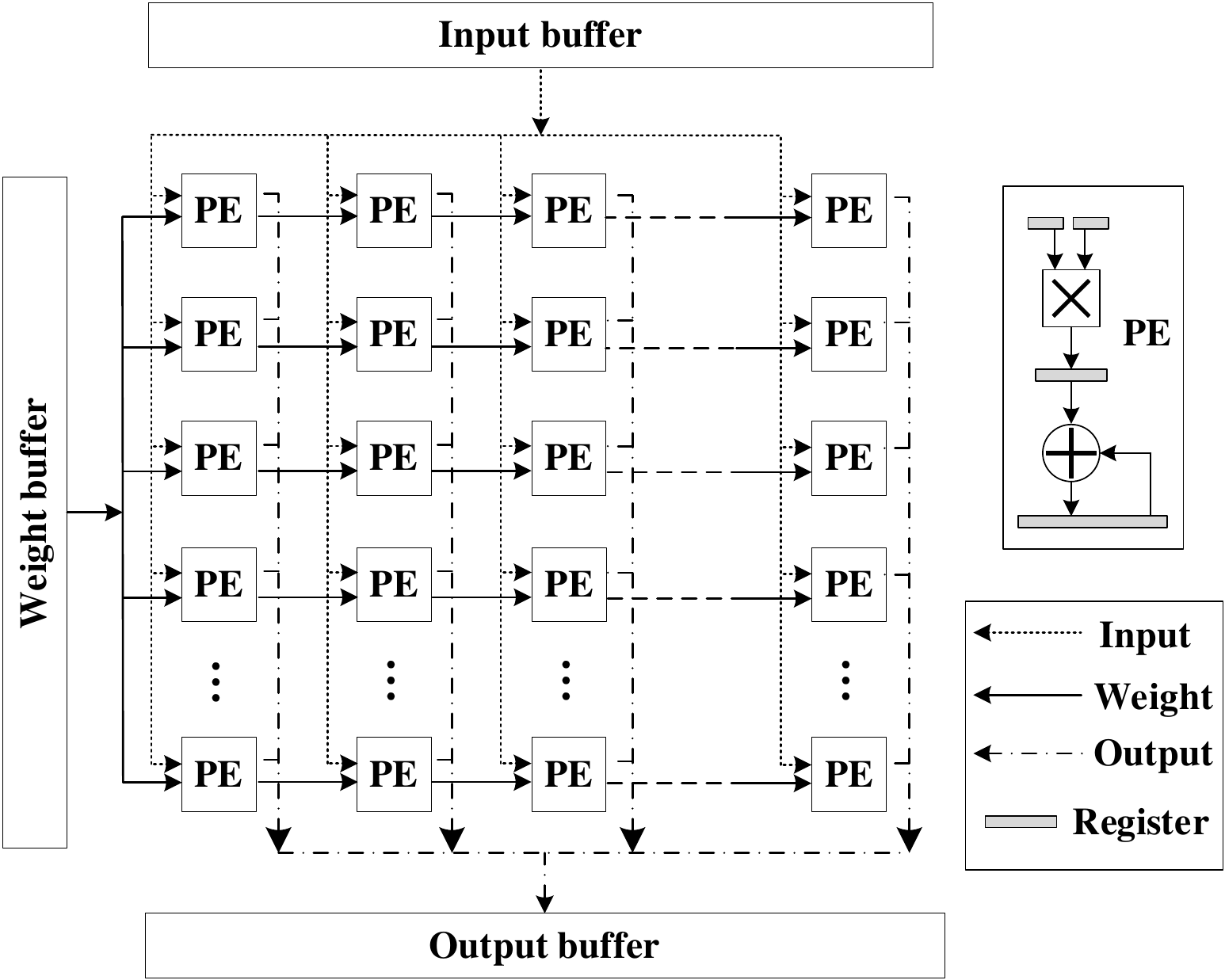}}
    \caption{A typical DLA with 2D computing array architecture.}
\label{fig:accelerator}
\vspace{-1em}
\end{figure}

A typical DLA with 2D computing array is shown in \autoref{fig:accelerator}. The computing array is composed of multiple homogeneous connected PEs. Each PE includes a multiplier and an accumulator. It only communicates with its four neighbors. Neural network operations such as convolution can be mapped to the computing array and executed in lock-step manner. To ensure a high-throughput neural network processing, the input features, weights and output features must be stored in the on-chip buffers to avoid external memory access stalls. In this work, we adopt a widely used output stationary dataflow for the neural network execution \cite{Chen2016Eyeriss}. The accumulation of each output feature stays stationary in a PE. The partial sum are stored in the same register file for accumulation to minimize the accumulation cost. In summary, each PE is responsible for the calculation of a single output feature and PEs in the same column calculate different output features in the same output channel. With the compact dataflow, the 2-D computing array can be fully utilized given limited on-chip buffer bandwidth provision when the neural network models are deployed on it.
\subsection{Deep Learning Accelerator Architecture}

\subsection{Influence of Faulty Computing Array}
To analyze the influence of hard errors on the above 2-D computing array, we inject random stuck-at bit errors to the registers of the PEs in a $32 \times 32$ computing array. We use bit error rate (BER) as the fault injection rate metric \cite{mittal2020survey}\cite{neggaz2018reliability}\cite{ares2018dac}, which refers to the total number of bit errors over the total bit number of the registers in the computing array. To facilitate the error characteristic of the 2-D computing array, we convert the BER to PE error rate (PER) instead. Both the input features and weights are 8-bit fixed point, so the input registers and the weight registers are set to be the same data width accordingly. The intermediate register in the PEs is set to be 16-bit and the accumulator in the PEs is set to be 32-bit in case of computing overflow. Thereby, there are 64 bit registers in total in each PE. While any persistent bit error in a PE is considered as an PE error, PER can be calculated using BER with (\autoref{eq:error_rate}). Basically, it means that the PE is correct only when none of the bit registers are wrong. Otherwise, the PE will be faulty.

\begin{small}
\begin{equation}
\label{eq:error_rate}
\begin{aligned}
    PER = 1-(1-BER)^{64}
\end{aligned}
\end{equation}
\vspace{-1em}
\end{small}

\begin{figure}
\setlength{\abovecaptionskip}{-1pt}
\setlength{\belowcaptionskip}{0pt}
	\center{\includegraphics[width=0.9\linewidth]{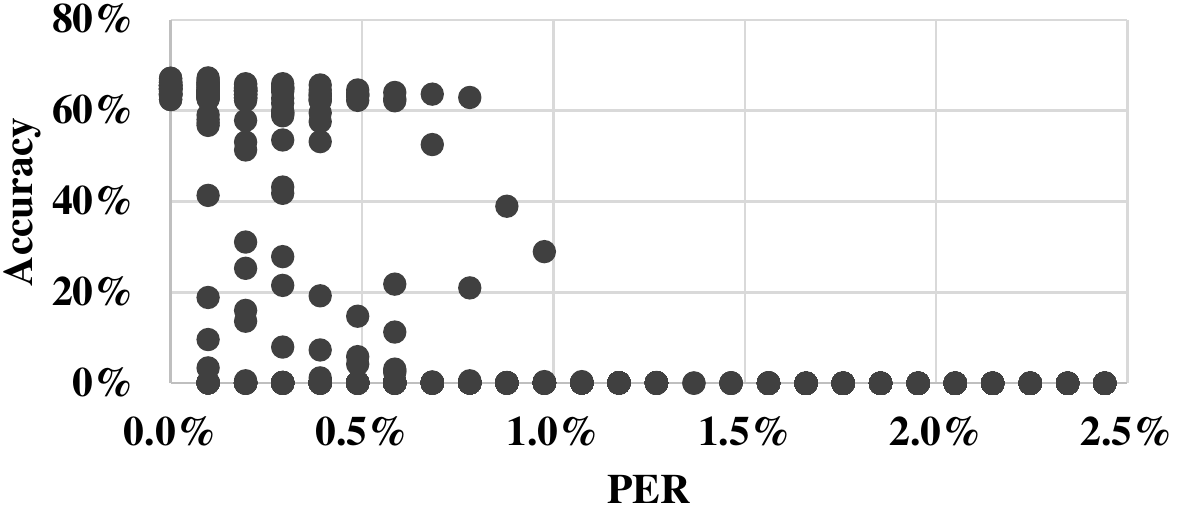}}
    \caption{Prediction accuracy of Resenet18 executed on a typical DLA under different PER setups. For each PER setup, 50 random fault configurations are evaluated on ImageNet.}
\label{fig:bit}
\end{figure}

We had random stuck-at bit errors injected to a DLA simulator implemented according to the architecture described in \autoref{fig:accelerator} for the fault analysis. We took Resnet18 pre-trained on ImageNet \cite{deng2009imagenet} as an example and had it implemented on the accelerator with random faults. In this case, we generated 50 random fault configurations and evaluated the prediction accuracy under different PER setups. The experiment result is shown in \autoref{fig:bit}. It reveals that the prediction accuracy varies dramatically across the different fault configurations. When the PER is higher than 1\%, the prediction accuracy of the model mostly degrades to zero. Moreover, we notice that the prediction accuracy may also drop considerably in some of the fault configurations even under very low PER. It indicates that the model accuracy depends on not only the PER but also the fault distribution. Thereby, protecting the computing array is required for mission-critical applications despite the fault injection rate. 

\begin{figure}
\setlength{\abovecaptionskip}{-1pt}
\setlength{\belowcaptionskip}{0pt}
	\center{\includegraphics[width=0.9\linewidth]{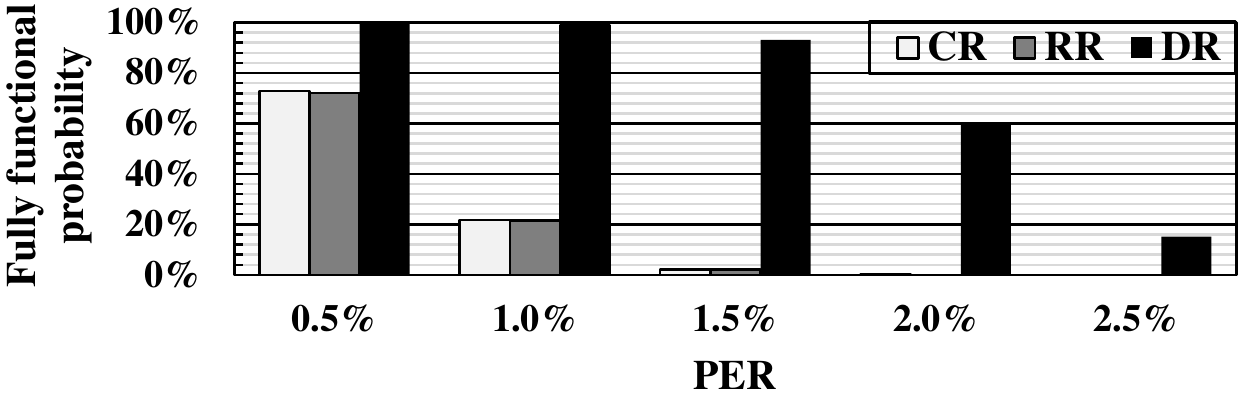}}
    \caption{The fully functional probability of the 2-D computing array under different PER setups.}
\label{fig:mo}
\vspace{-1em}
\end{figure}

In addition, we further evaluated the classical hardware redundancy strategies, i.e. RR, CR, and DR for the regular 2-D computing array and measured the fully functional probability of the computing array under different PER setups. The evaluation result is shown in \autoref{fig:mo}. It can be observed that these classical redundant design approaches can hardly mitigate all the faulty PEs even when the PER is around 1\% which indicates that there are only 10 faulty PEs on average. In contrast, the number of redundant PEs is 32, which is much larger than the number of faulty PEs. It demonstrates that the redundant PEs cannot be fully utilized by these redundancy strategies because of the unevenly faulty PE distribution. The situation further deteriorates with the increase of the PER, which can be rather risky for the mission-critical AI applications.

\section{HyCA for Fault-tolerant DLAs} \label{sec:overview}
In this section, we will present an overview of the proposed hybrid computing architecture (HyCA) for fault-tolerant DLAs first. Then, we will illustrate the dataflow for the fault mitigation, HyCA microarchitecture, and fault detection with HyCA respectively.

\subsection{HyCA Overall Architecture}
\begin{figure}
\setlength{\abovecaptionskip}{-2pt}
\setlength{\belowcaptionskip}{0pt}
	\center{\includegraphics[width=0.9\linewidth]{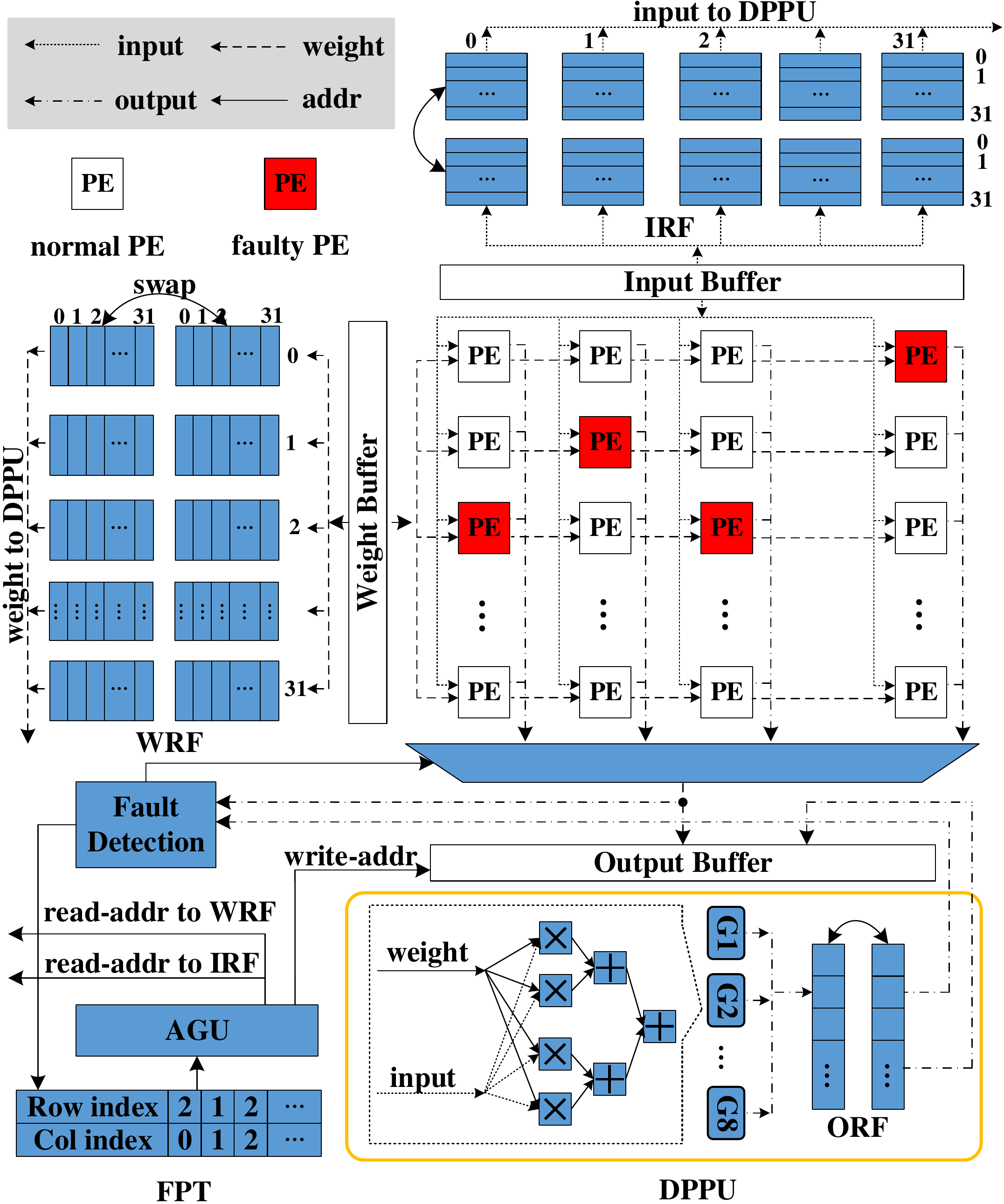}}
    \caption{Overview of of a DLA with Hybrid Computing Architecture. The components highlighted with blue are added to the conventional DLA to tolerate faulty PEs in arbitrary locations of the 2-D computing array.}
\label{fig:mapping}
\vspace{-1.5em}
\end{figure}

In order to tolerate various fault configurations with a unified computing architecture, we propose a hybrid computing architecture (HyCA), which has a dot-production processing unit (DPPU) seated along with the regular 2-D computing computing array, to recompute all the operations mapped to the faulty PEs in arbitrary locations of the computing array as shown in \autoref{fig:mapping}. While the 2-D computing array has each PE to calculate the different output features sequentially given the output stationary dataflow \cite{Chen2016Eyeriss} and the DPPU has all the PEs to compute a single output features in parallel, they have distinct read patterns of the input features and weights from the corresponding on-chip buffers. More specifically, the 2-D computing array needs to read an array of input features in the same row and channel in each cycle while DPPU needs to read an array of input features aligned in channel dimension in each cycle. Thereby, the on-chip buffers cannot fulfill the read operations of the two computing units at the same time due to the limited read ports and distinct data layout requirements. To make sure that the normal 2-D array processing will not be affected by the DPPU recomputing, the on-chip buffer design remains unchanged. In this case, DPPU cannot read the required weights and input features aligned in channel dimension if it starts the recomputing at the same time with the 2-D computing array. 

\begin{figure*}[tb]
\setlength{\belowcaptionskip}{0pt}
	\center{\includegraphics[width=0.95\linewidth]{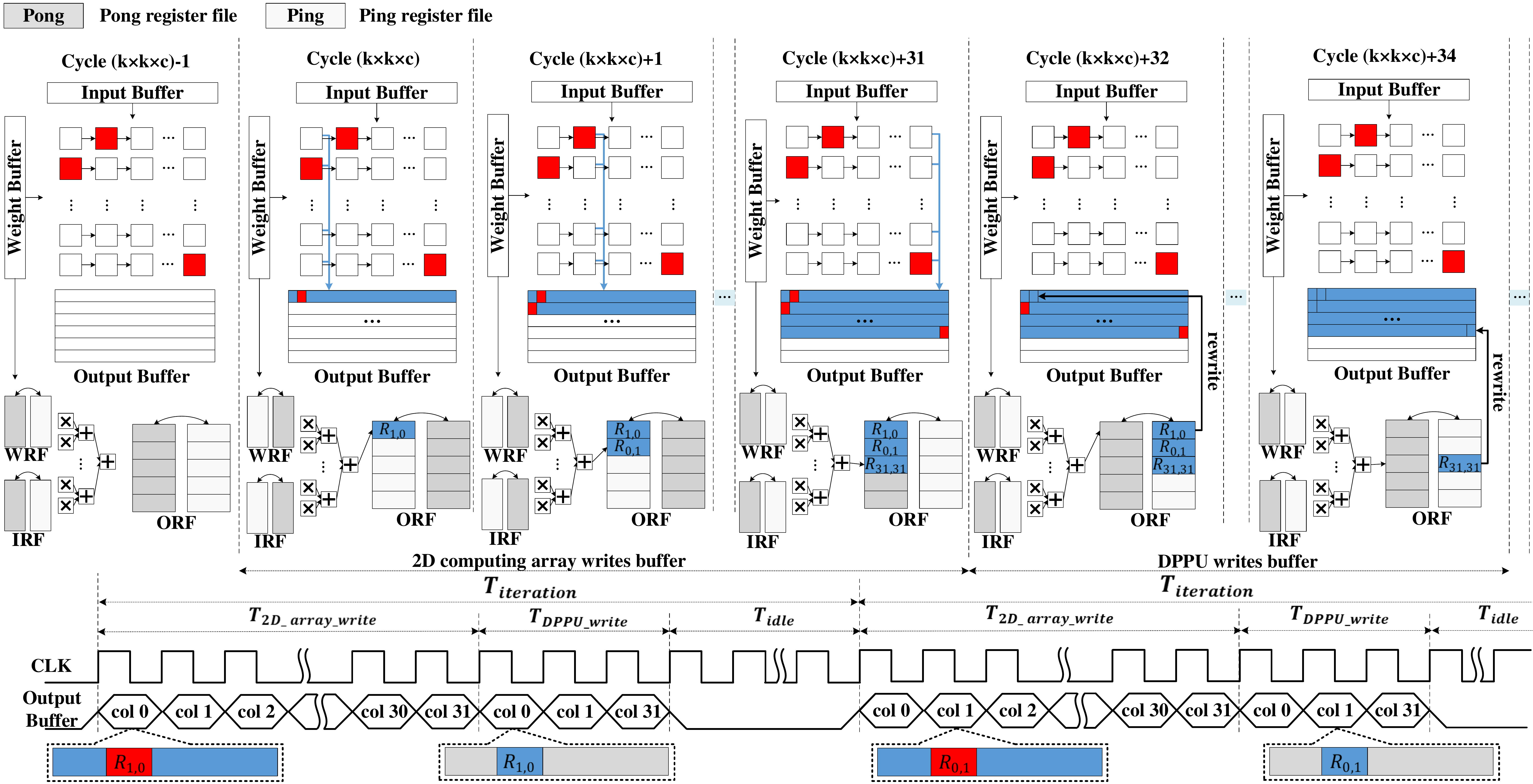}}
    \caption{The dataflow of using DPPU for recomputing the neural network operations mapped on the faulty PEs of a typical DLA. It also demonstrates how the DPPU overwrites the faulty computing results produced by the 2-D computing array.}
\label{fig:pipeline}
\vspace{-1em}
\end{figure*}

To address the problem, we have the input features and weights buffered in an input register file (IRF) and a weight register file (WRF) respectively while they are read for the 2-D computing array processing. Meanwhile, we have the recomputing delayed until there are sufficient inputs and weights ready for the recomputing. Accordingly, the delay must be larger than or equal to the number of weights required by DPPU data consumption in a single cycle to ensure DPPU can be fully utilized. As the DPPU may recompute operations on any PE in the 2-D computing array, the delay also needs to be larger than or equal to $Col$ when the last column of the PEs obtain the weights passed from the first column of PEs. Suppose $D$ represents the delay, then $D \geq Col$. Note that $Col$ refers to the column size of the 2-D computing array. In this work, we organize IRF and WRF in Ping-Pong manner to ensure that the 2-D computing array can continue the normal dataflow without any stall during the DPPU recomputing. As the DPPU conducts the output feature calculation in parallel, DPPU can always finish the recomputing of the operations mapped to the faulty PEs before the Ping-Pong register files swap with each other when the DPPU size does not exceed the number of the faulty PEs. Note that DPPU size refers to the number of multipliers in DPPU. Since the peak computing power of DPPU equals to that of the 2-D computing array when configured with the same number of PEs, DPPU size is comparable to the 2-D computing array size and can also be used to represent its computing power. This also explains why DPPU can always finish the recomputing tasks before new weights and inputs are ready when DPPU size is larger than the number of faulty PEs in the 2-D computing array. 

In addition, we have a fault PE table (FPT) to record the coordinates of the faulty PEs in the 2-D computing array which can be usually obtained with a power-on self-test procedure. With the coordinates, an address generation unit (AGU) is used to generate the read addresses and instruct the DPPU to read the right input features and weights from the register files. Moreover, AGU also determines the addresses to the output buffer for the overlapped writes of the recomputed output features. Similar to the IRF and WRF, there is also a Ping-Pong register file called ORF for the DPPU outputs and it is utilized to pipeline the DPPU recomputing and the write from DPPU to the output buffer.

While DPPU can be utilized to calculate any output features mapped to the 2-D computing array, we can also use DPPU to check whether the calculation of an output feature in the 2-D computing array is correct, which can be used to detect the wear-out or aging induced persistent errors at runtime. If the computing results obtained from the 2-D computing array and DPPU do not match, it indicates that the corresponding PE in the 2-D computing array is faulty as DPPU with much smaller sizes compared to the 2-D computing array can be easily protected with much less overhead and is usually considered to be correct. By changing the fault PE table and scanning the computing of all the PEs in the 2-D computing array sequentially, we can detect the PE faults at runtime without affecting the 2-D computing array processing. Basically, the redundant recomputing mechanism can be mostly reused by the fault detection. And we only need a tiny fault detection module to conduct the scanning of the 2-D computing array and the comparison to the DPPU processing. Details of the fault detection module will be illustrated in the rest of this section.


\subsection{HyCA Dataflow for Fault Mitigation} \label{sec:dataflow}
To further illustrate the dataflow in HyCA especially the redundant computing unit DPPU, we take HyCA with a $32 \times 32$ 2-D computing array and three faulty PEs as an example. DPPU in the HyCA has 32 PEs included. The example is shown in \autoref{fig:pipeline}. Suppose $c$ and $k$ represent the input channel depth and the convolution kernel size respectively. It takes a PE in the 2-D computing array $k \times k \times c$ cycles to produce a convolution output. Without loss of generality, assume that the example starts at Cycle $k \times k \times c - 1$ when the first column of PEs complete a set of output feature calculation. The 2-D computing array occupies the output buffer until the last column of PEs complete the output feature data. Afterwards, the 2-D computing array may start to compute the new output features, but it usually takes $k \times k \times k \times c$ cycles to complete with the output stationary dataflow. In this case, DPPU starts to use the output buffer and update the recomputed results to the output buffer without write conflicts. The processing steps are detailed as follows.

\begin{enumerate}
    \item At cycle $k\times k\times c - 1$, the first column of PEs produce a column of output features accordingly and pass the weights to the second column of PEs. At the same time, the weights and input features used in the first column of PEs are stored in the corresponding Pong register file.
    
    \item At cycle $ k\times k\times c $, the first column of PEs write the calculated output features to the output buffer and start the calculation of new output features. Since PE(1,0) is faulty, the computing result written to the output buffer from this PE as marked with red color was wrong. While the input features and weights that are used for the output feature calculation on PE(1, 0) remain stored in the IRF and the WRF respectively, they will be read to the DPPU for the recomputing at this cycle. 
    
     \item At cycle $ k\times k\times c + 31$, the Pong WRF and the Pong IRF are filled with the newly incoming weights and inputs, and they will be kept for 32 cycles. Weights and input features coming in next cycle will overwrite the data in the Ping WRF and the Ping IRF respectively. Thereby, DPPU must finish the recomputing that depends on the weights stored in the Ping WRF and IRF at this cycle. Otherwise, the data in the Ping register files will be overwritten. Afterwards, the processing repeats from the first processing step for another 32 cycles until the end of the convolution calculation.
    
    \item At cycle $ k\times k\times c + 32$, DPPU has the recomputed output features in the output register file (ORF) written to the output buffer with a byte mask such that only the recomputed output feature is updated. Meanwhile, it starts to recompute the latest set of output features that are mapped to the faulty PEs in the 2-D computing array. As each output feature calculation is mapped to a single PE in the 2-D computing array using the classical output stationary dataflow, it takes $c \times k \times k$ cycles to complete an output feature calculation. As $c \times k \times k$ is usually larger than 32, and the output buffer will be occupied for only 32 cycles during each set of output feature calculation, the recomputed output feature data can be updated to the output feature buffer without conflicts. 

    \item At cycle $ k\times k\times c + 34$, because there are only three faulty PEs in the 2-D computing array, it takes the DPPU three cycles to have the recomputed results overwritten from the ORF to the output buffer.
    
    \item From cycle $ k\times k\times c + 35$ to $ k\times k\times c + c$, both the 2-D computing array and the DPPU conduct the partial convolution locally, so the output buffer port is idle before the first column of PEs complete the new output feature calculation. 

\end{enumerate}

As shown in \autoref{fig:pipeline}, the overall processing is conducted iteratively. Each iteration includes a set of complete output feature calculation and it can be divided into three phases, i.e. 2-D array write, DPPU write, and idle from the perspective of the output buffer status. While each PE produces an output feature data per iteration and a PE conducts one MAC per cycle, the processing time of an iteration is $T_{iteration}=c \times k \times k$. For the 2-D computing array write, it takes $T_{2D\_arrary\_write} = D$ cycles per iteration where $D$ refers to the number of cycles that DPPU delays after the 2-D computing array processing. For the DPPU write, it needs $T_{DPPU\_write}=fault\_PE\_num$ cycles as the DPPU updates the recomputed output features sequentially. However, DPPU recomputing does not have to start after the entire computing of the output features on the 2-D computing array. Instead, it is pipelined with the 2-D computing array but only $D$ cycles slower. Thus, the weights and the input features consumed by the 2-D computing array must be fully accommodated by the Ping-Pong register files during the $D$ cycles. Accordingly, the depth of weight and input feature Ping-Pong register files is set to be $2\times D \times Row$. To minimize the register file overhead, we set $D=Col$.

As the average throughput of a PE in the 2-D computing array is the same with that in the DPPU, each multiplier in the DPPU can be used to repair a faulty PE in the 2-D computing array. Thereby, the DPPU size essentially represents the capability of the fault tolerance of the proposed HyCA without performance penalty. When the number of faulty PEs is larger than the DPPU size, we seek to preserve the computing power as much as possible without altering the target neural network models. To that end, we discard the faulty PEs that cannot be repaired due to the lack of the computing redundancy in the DPPU. As it is usually inefficient to compile and deploy the neural network models to a computing array with irregular row sizes which can cause both the irregular on-chip buffer accesses and external memory accesses, we choose to discard the columns with unrepaired faulty PEs and the columns that are disconnected from the input/weight/output buffers. Moreover, HyCA can repair any faulty PEs in the 2-D computing array, so it offers more flexibility to prioritize the faulty PEs for repairing such that the surviving computing array can be maximized especially when there are insufficient redundant PEs. In this work, the maximum remaining computing array can be obtained simply by assigning higher repairing priority to the faulty PEs on the left, which ensures that the surviving computing array is connected to the on-chip buffers. 

\subsection{HyCA Microarchitecture} \label{remaining-array}
In this section, we will illustrate the major components of HyCA added to the baseline DLA and they include the DPPU, register files, and the fault PE table (FPT). FPT keeps the coordinates of the faulty PEs that will be repaired by the DPPU. As the maximum number of faulty PEs that can be tolerated without performance penalty is determined by the DPPU size, FPT is configured with $DPU\_size$ entries accordingly. Address generation unit (AGU) is a piece of control logic that generates the access addresses of the weight register file, input register file, and output register file based on the FPT for the recomputing of the DPPU. The structures of FPT and AGU are simple and we will not dwell on it. In contrast, the DPPU and the register files are relatively more complex, and they dominate the hardware overhead. Thus, they will be detailed in the rest of this section.

\begin{figure}
\setlength{\abovecaptionskip}{-10pt}
\setlength{\belowcaptionskip}{0pt}
	\center{\includegraphics[width=1\linewidth]{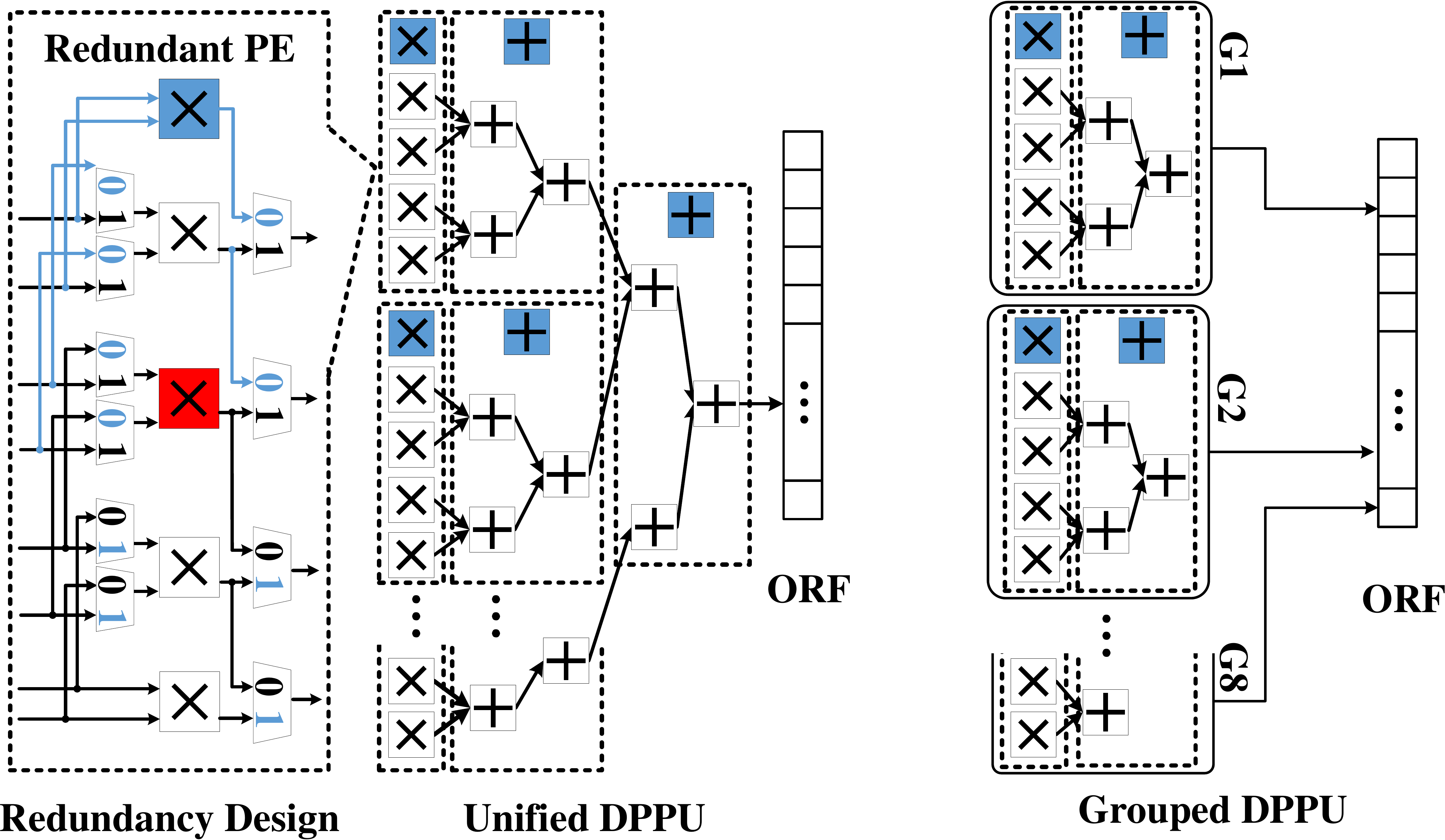}}
    \caption{Structures of the unified DPPU and the grouped DPPU. For both the unified DPPU and the grouped DPPU, they are protected with redundant PEs. Each redundant PE is used to protect a set of homogeneous PEs and these PEs are connected with ring topology to reduce the signal fan-out.}
\label{fig:deg}
\vspace{-1em}
\end{figure}

\subsubsection{Dot-Production Processing Unit (DPPU)}
DPPU is utilized for the dot-production and it consists of a set of multipliers as well as an adder tree that is used to aggregate all the multiplication results in a pipelined manner. It is mainly used to recompute the output features that are mapped to the faulty PEs in the 2-D computing array. An intuitive implementation is to construct a single unified dot-production unit which has both the input features and weights read from the corresponding register files in a single cycle and processed in parallel. As DPPU starts $Col$ cycles later after the 2-D computing array, each faulty PE in the 2-D computing array has $Col$ weights and input features multiplied and accumulated. Accordingly, $Col$ weights and input features can be extracted for the recomputing on DPPU for each faulty PE in the 2-D computing array. In order to make best use of the PEs in the DPPU, the $Col$ weights and input features must be fully distributed to the DPPU. If the entire DPPU is organized into a unified dot-production unit, the size of the DPPU is rather limited, which hinders the scalability of the DPPU. For instance, when $Col$ is set to be 32 and DPPU size is set to be 24 or 48, the computing mapped to a single faulty PE cannot be perfectly mapped to the DPPU, which will lead to the under-utilization of the DPPU. To address this problem, we propose to divide the PEs in the DPPU into multiple smaller groups and each group can conduct the dot-production independently. As the number of PEs in each group gets smaller, they are more likely to be fully utilized by the computing mapped to a faulty PE. As shown in \autoref{fig:deg}, each group includes 8 PEs and it completes the computing of a faulty PE in 4 cycles when $Col$ is set to be 32. In this case, the DPPU size can be scaled conveniently. At the same time, the different groups can conduct operations mapped to different faulty PEs in the 2-D computing array in parallel.  

While the DPPU is used to recompute all the faulty operations in the 2-D computing array, it must be resilient enough to ensure the functionality. Otherwise, a single fault in the DPPU may corrupt the whole accelerator. To improve the resilience of the DPPU, we add redundant PEs to the DPPU as shown in \autoref{fig:deg}. Basically, the multipliers used in the DPPU are divided into groups and each group is equipped with a redundant multiplier. Instead of having the redundant multiplier shared by all the multipliers in the group, we have the redundant multiplier and the multipliers in the group connected in a directed ring topology and each multiplier can be configured to replace its downstream neighboring multiplier. When any of the multiplier fails, it can be replaced by its upstream multiplier immediately. Compared to the shared redundancy design, this approach can avoid high fan-out connections to the redundant multipliers. Similarly, we also have the adders in the adder tree protected with the same redundancy design approach. 

\begin{figure}
\setlength{\abovecaptionskip}{0pt}
\setlength{\belowcaptionskip}{0pt}
	\center{\includegraphics[width=0.95\linewidth]{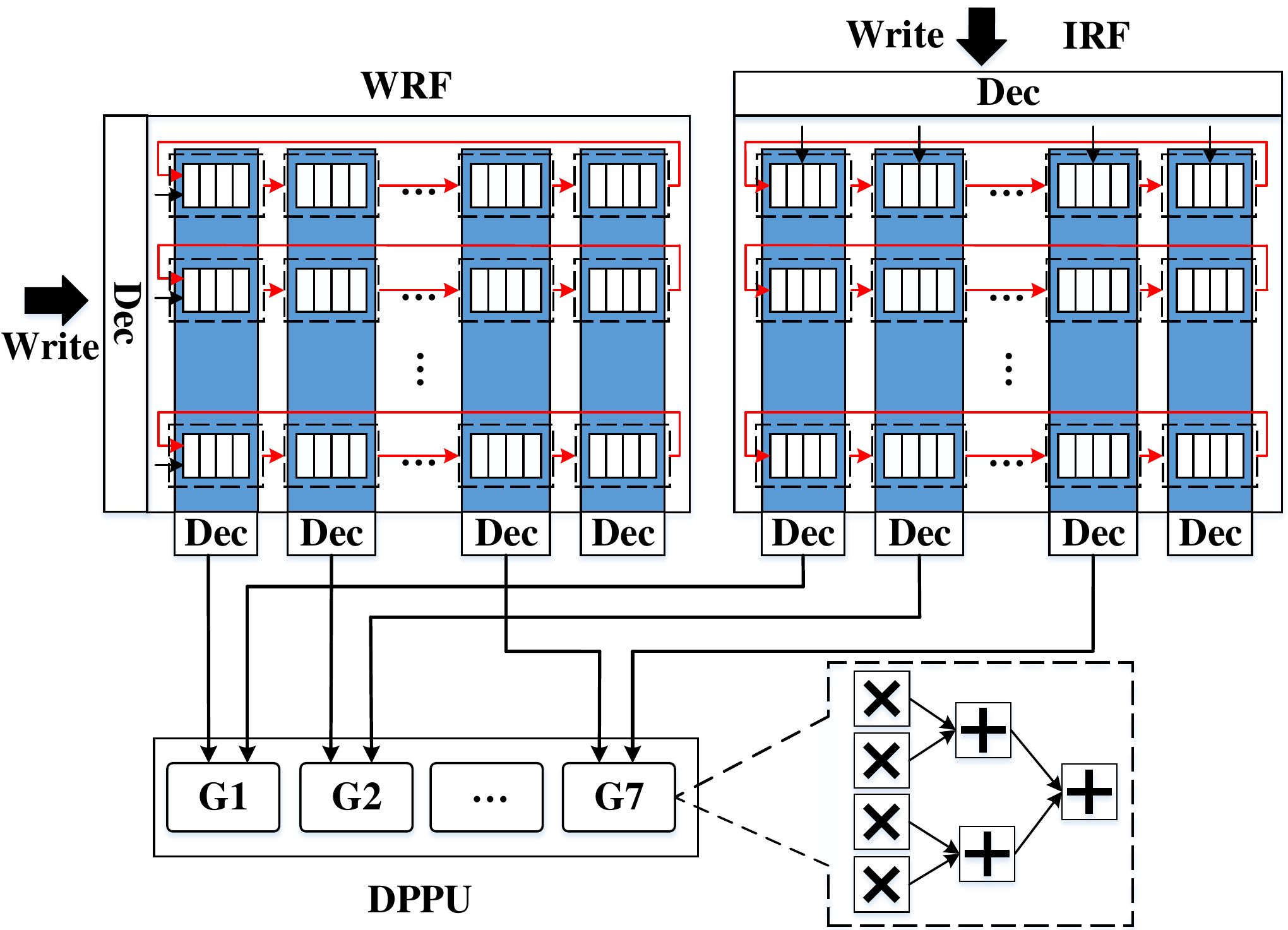}}
    \caption{Organization of the Weight Register File (WRF) and Input Register File (IRF). It shows how the register files are connected with a grouped DPPU with different number of computing groups.}
\label{fig:reg}
\vspace{-1em}
\end{figure}

\subsubsection{Register Files}
The IRF and the WRF are used to back up the data read from input buffer and weight buffer, and then supply the data to the DPPU for the recomputing. As the 2-D computing array and the DPPU have different dataflows for the neural network computing, the weight register file is written in column-wise manner but read in row-wise manner as shown in Figure \ref{fig:reg}. When the DPPU is split into multiple groups, these groups will be responsible for different faulty convolution operations and they need to read different rows of WRF and IRF at the same time. Although a straightforward multi-port register file can fulfill the concurrent register file read, it will induce substantial hardware overhead according to \cite{Energy2003Aneesh,A2012Chang}. 

While we observe that each computing group in DPPU has only a small number of PEs and they cannot consume a single row of inputs and weights in a single cycle. As a result, the straightforward multi-port register file actually has the bandwidth wasted. With this observation, we also have the register files split into groups in row direction such that each group of the register file can be read independently by the corresponding computing group in the DPPU. In this case, each register file has only a single read port, but each computing group can only read a segment of the data in the register files as indicated by the blue color. While an output feature recomputing on DPPU needs an entire row of data in the register files, we have each row of the registers organized as a circular shift register. With the shift register, different segments of the data in a row can be obtained by the corresponding computing group in DPPU in a few cycles. At the same time, the amount of data fed to each computing group can be fully utilized. Moreover, we notice the read port data width of the register files is not necessarily equal to the DPPU size. When the DPPU size is larger than the register file data width, more read ports can be added to some of the register file groups rather than the entire register file. When the DPPU size is smaller than the register file data width, some of the register file groups do not even need a read port as shown in Figure \ref{fig:reg}. Thereby, the DPPU size can be scaled conveniently and it will not be limited by the register file sizes. 

\subsection{Fault Detection with HyCA}
\begin{figure}[tb]
\setlength{\abovecaptionskip}{-1pt}
\setlength{\belowcaptionskip}{-10pt}
	\center{\includegraphics[width=0.95\linewidth]{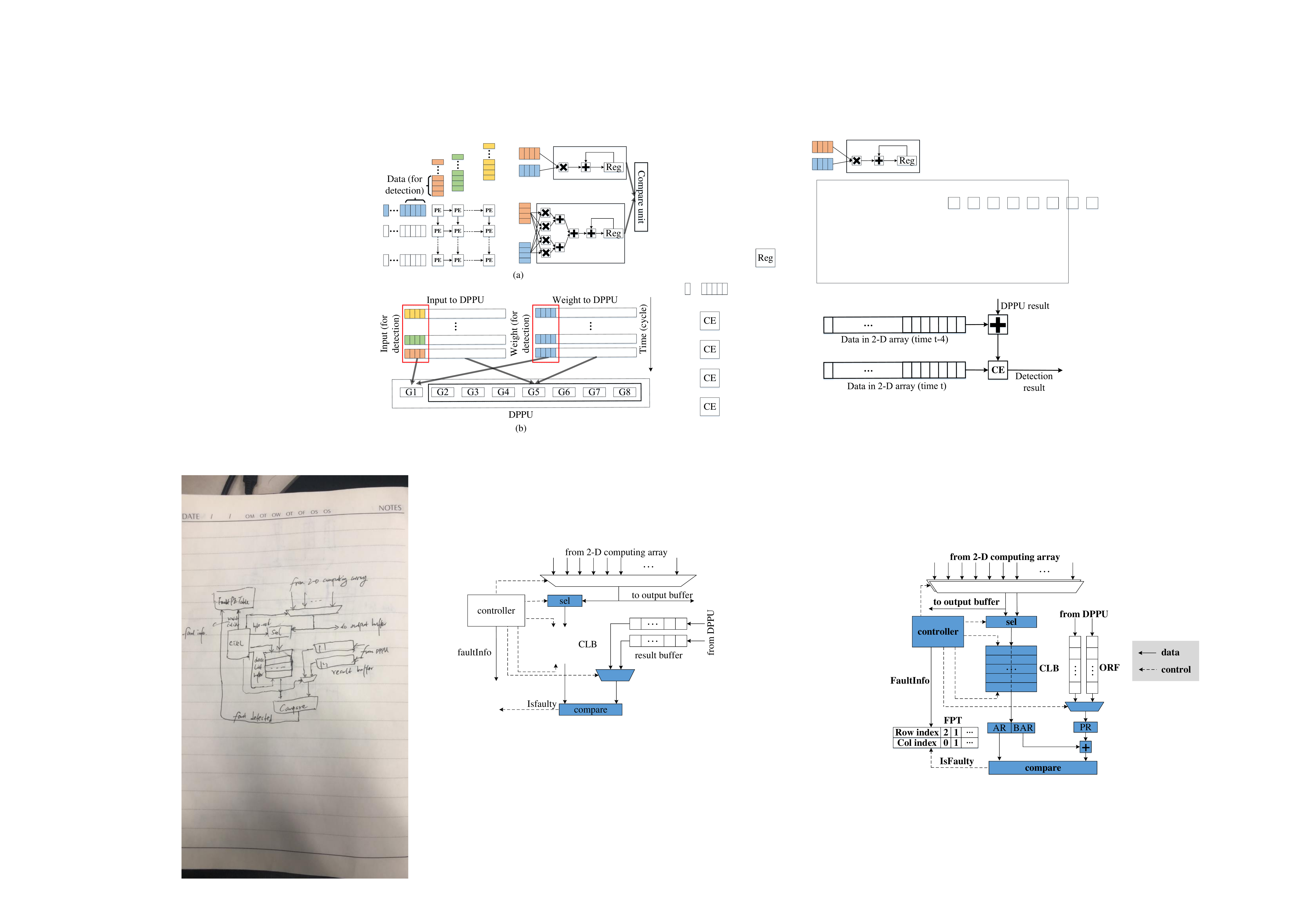}}
    \caption{Structure of the Fault-Detection Module.}
\label{fig:detection}
\vspace{-1.5em}
\end{figure}

On top of the fault mitigation, DPPU can also be utilized to conduct fault detection at runtime. The basic idea is to have the DPPU to recompute the operations on a PE in the 2-D computing array. Then, we have the computing results compared to check if the PE in the 2-D computing array is faulty. By scanning all the PEs in the 2-D computing array sequentially, we can determine if the 2-D computing array is faulty. While the DPPU always starts the recomputing $Col$ cycles later, the computing result of a PE to be checked is already updated or written to the output buffer when the DPPU completes the recomputing. To address the problem, we have the computing results to be checked buffered in a checking list buffer (CLB) as shown in Figure \ref{fig:detection}. As the fault detection scanning is conducted sequentially, a simple on-chip buffer can fulfill the requirements. When the DPPU completes the recomputing, the fault detection module can have the results compared with that stored in the CLB.  

As hard faults in a PE can usually lead to computing errors of most of the computation, we do not have to compare the final output feature computing results for the fault detection. Instead, we can compare the partial computing results of a PE for the fault detection such that the fault detection can be faster and more efficient. Since the DPPU conducts the processing in parallel, we have the partial computing result (PR) produced by a single DPPU group in a single cycle compared in this work. Different from the DPPU, PEs in the 2-D computing array have the computing results accumulated continuously before the entire output feature processing is completed. Suppose only one DPPU group is reserved for the runtime fault detection and the DPPU group includes $S$ PEs. To enable the comparison for the fault detection, we have both the base accumulated results (BAR) and the accumulated results (AR) calculated $S$ cycles later stored in the CLB. In the next cycle, another pair of the BAR and the AR from different PEs will be stored accordingly. As the weights and input features are stored in their register files for only $Col$ cycles, we only have $Col$ ARs and BARs stored in the CLB. While the CLB is also organized in Ping-Pong manner, the total size of the CLB is $4 \times W \times Col$ Bytes where $W$ denotes the width of the accumulator in PEs. 

According to the recomputing dataflow, the reserved DPPU group performs the recomputing $Col$ cycles later. Unlike the fault-tolerance oriented recomputing, the fault-detection oriented recomputing only conducts the dot-production of $S$ weights and input features rather than $Col$ weights and input features. The results of the DPPU will be compared with the corresponding results stored in the CLB for the fault detection. Basically, AR will be compared to the addition of PR and BAR. When a faulty PE is detected, the faulty information i.e. the fault PE row index and column index will be updated to the FPT. One comparison can be done per cycle, so it takes the fault detection module $Col$ cycles to complete the comparison with the stored results in CLB. Accordingly, it takes the fault detection module $Row \times Col + Col$ cycles to complete the fault detection of the whole 2-D computing array. When a DPPU group has more PEs included, it needs to check a partial result with more computation. As a result, the fault detection time is independent with the number of PEs in the DPPU group. While the fault detection time is already much smaller than the processing time of a normal neural network layer, it is fast enough to detect the computing errors. To avoid the frequent fault detection, the fault detection module can be activated periodically in a larger time range depending on the requirements of the applications. In addition, the fault detection module can reuse the majority of the fault recovery design, it induces only some simple controlling logic and a small CLB, and consumes negligible chip area.

\section{Experiment} \label{sec:evaluation}
\subsection{Experiment Setup}
\subsubsection{Accelerator Configurations} The proposed deep learning accelerator with HyCA is implemented in Verilog and synthesized with Design Compiler under TSMC 40nm technology. The computing array size and DPPU size are set to be $32\times 32$ and 32 respectively. The input feature buffer size is 128KB, output feature buffer size is 128KB and the weight buffer size is 512KB. The computing of DPPU is delayed by $D=32$ cycles after the 2-D computing array, so both the weight register file size and the input register file size are set to be $2 \times 32 \times D = 2048$ i.e. 2KB. The output register file in DPPU is 64-byte. The fault PE table size is $32 \times 10$bits. Each entry of the table includes 5-bit row index and 5-bit column index of a faulty PE. Both the data width of weights and input feature data is 8-bit. To ensure the resilience of the DPPU, we have every four multipliers in the DPPU grouped and equipped with a redundant multiplier, and every three adders in the DPPU grouped and protected with a redundant adder. For the 2-D computing array, we have three classical redundancy approaches including RR, CR, and DR implemented and each redundancy implementation is equipped with 32 redundant PEs.

\subsubsection{Fault Models} To evaluate the reliability of the redundancy designs comprehensively, we have two different fault distribution models including the random distribution model and the clustered distribution model implemented. For the random distribution model, the faults are randomly distributed across the entire computing array. For the clustered distribution model which is usually used to characterize the manufacture defects, the faults are more likely to be close to each other and the model proposed in \cite{meyer1989modeling} is applied in this work. Meanwhile, we notice that the influence of hardware faults is related with the fault distribution, so we generate 10000 configurations randomly for each fault injection rate and average the evaluation in the experiments. 

As hard errors are mainly caused by the manufacturing defects, aging, and wear-out, which can be affected by many complex factors such as application requirements, working status, and the fabrication, there is still a lack of references investigating the practical PER setup and prior fault-tolerant designs typically have distinct error rate setups \cite{liu2011resilient, li2008understanding, abdullah2020salvagednn, analyzing2018vts}. In this case, we evaluate the hard error rate in a large range and seek to demonstrate when we can ensure reliable computing. Then, we expect the users to choose the target hard error rate for their specific fault-tolerant designs. In addition, since we mainly focus on the reliability of the regular 2-D computing array in a deep learning accelerator, we use PE error rate (PER) as the fault injection metric similar to the works in \cite{zhang2018analyzing} and \cite{qian2016optimal}. Meanwhile, we notice that bit error rate (BER) that refers to the number of bit errors over the total number of memory bits is directly related with definition of the hard error on chips, and it has been widely utilized as a metric for fault analysis in many prior works \cite{mittal2020survey}\cite{neggaz2018reliability}\cite{ares2018dac}. Thus, we convert the BER to PER with (\autoref{eq:error_rate}) mentioned in Section \ref{sec:motivation} assuming that any bit error in a PE will cause the PE failure. While BER typically ranges from $1 \times 10^{-7}$ to $1 \times 10^{-3}$, PER ranges from 0$\%$ to 6$\%$ according to the conversion.

\subsubsection{Neural Network Benchmark} To evaluate the performance of a typical DLA with the proposed fault-tolerant HyCA, we have a set of representative neural network models including Alexnet, VGG, Resnet, and YOLO used as the benchmark. Alexnet, VGG and Resnet are classical models used for image classification, while YOLO is mostly used for object detection. All the models are pre-trained on ImageNet. We measured the performance of the benchmark on the DLAs with different redundancy design approaches using Scale-sim\cite{samajdar2018scale}. Since Scale-sim is relatively slow, it is difficult to obtain the performance of all the random fault configurations directly. In this experiment, we determined the final valid computing array setups of all the fault configurations and performed the simulation on only the unique computing array setups. As many fault configurations lead to the same computing array setups eventually, this approach greatly reduces the simulation time. Finally, we averaged the resulting performance based on the generated fault configurations.

\begin{figure}
\setlength{\abovecaptionskip}{-2pt}
\setlength{\belowcaptionskip}{0pt}
	\center{\includegraphics[width=0.9\linewidth]{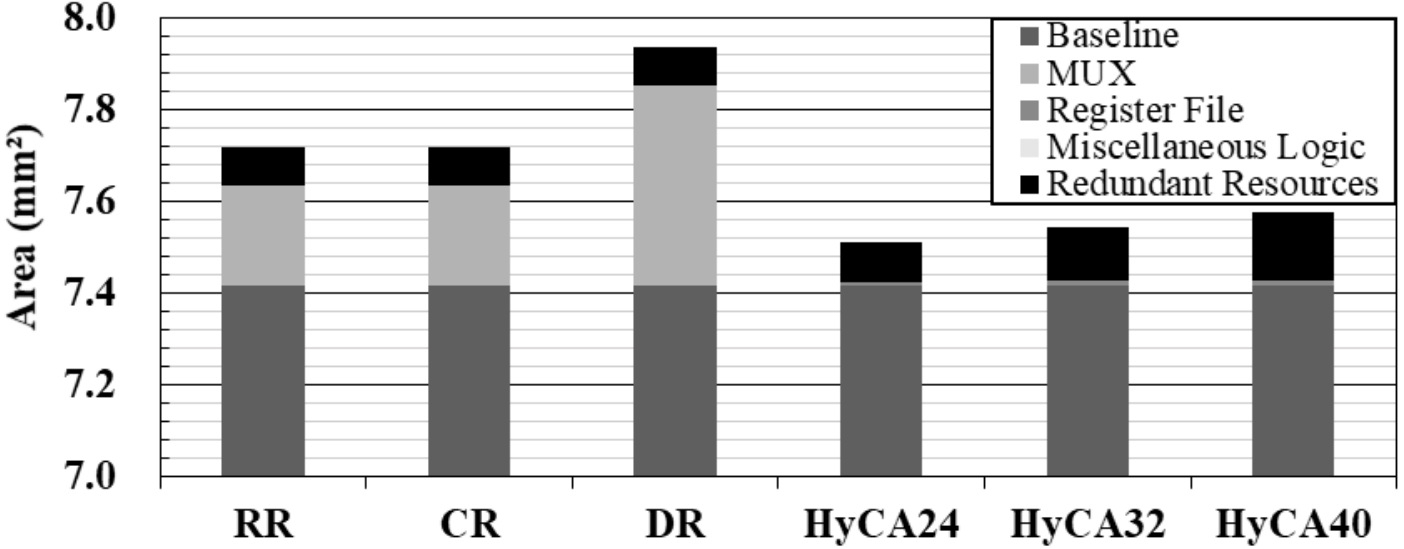}}
    \caption{Chip area under different redundancy approaches.}
\label{Circuit area}
\vspace{-1em}
\end{figure}

\subsection{Chip Area Overhead Comparison} \label{sec:chip-area}
\autoref{Circuit area} illustrates the chip area of the DLAs with different redundancy approaches including RR, CR, DR, and HyCA. Particularly, we have three HyCA-based designs with different DPPU sizes compared. The DPPU size of HyCA24, HyCA32, and HyCA40 is 24, 32 and 40 respectively. According to the comparison \autoref{Circuit area}, the HyCA-based designs exhibit much less redundancy overhead compared to the classical redundancy designs. The redundancy overhead of the HyCA-based designs mainly consist of the redundant PEs and the register files, while the redundancy overhead of the RR-based, CR-based, and DR-based designs are mainly attributed to the MUX and the redundancy PEs. As the 2-D computing array size is $32 \times 32$, the number of redundant PEs in RR-based, CR-based, and DR-based designs is the same and the chip area caused by the redundant PEs is also equal. While HyCA has different PE structures, i.e. independent multipliers and adders rather than MACs, and additional redundant PEs, the chip area of the redundant PEs is larger given the same DPPU sizes. In contrast to the chip area of the redundant PEs in HyCA, the added small Ping-Pong register files in HyCA consumes much less chip area. Different from HyCA, RR, CR and DR include a large number of MUX to enable the replacement of faulty PEs with the redundant PEs. These MUX take up substantial chip area and dominate the redundancy overhead. 

\subsection{Reliability Comparison}
\begin{figure}
\setlength{\abovecaptionskip}{-10pt}
\setlength{\belowcaptionskip}{-2pt}
	\center{\includegraphics[width=1\linewidth]{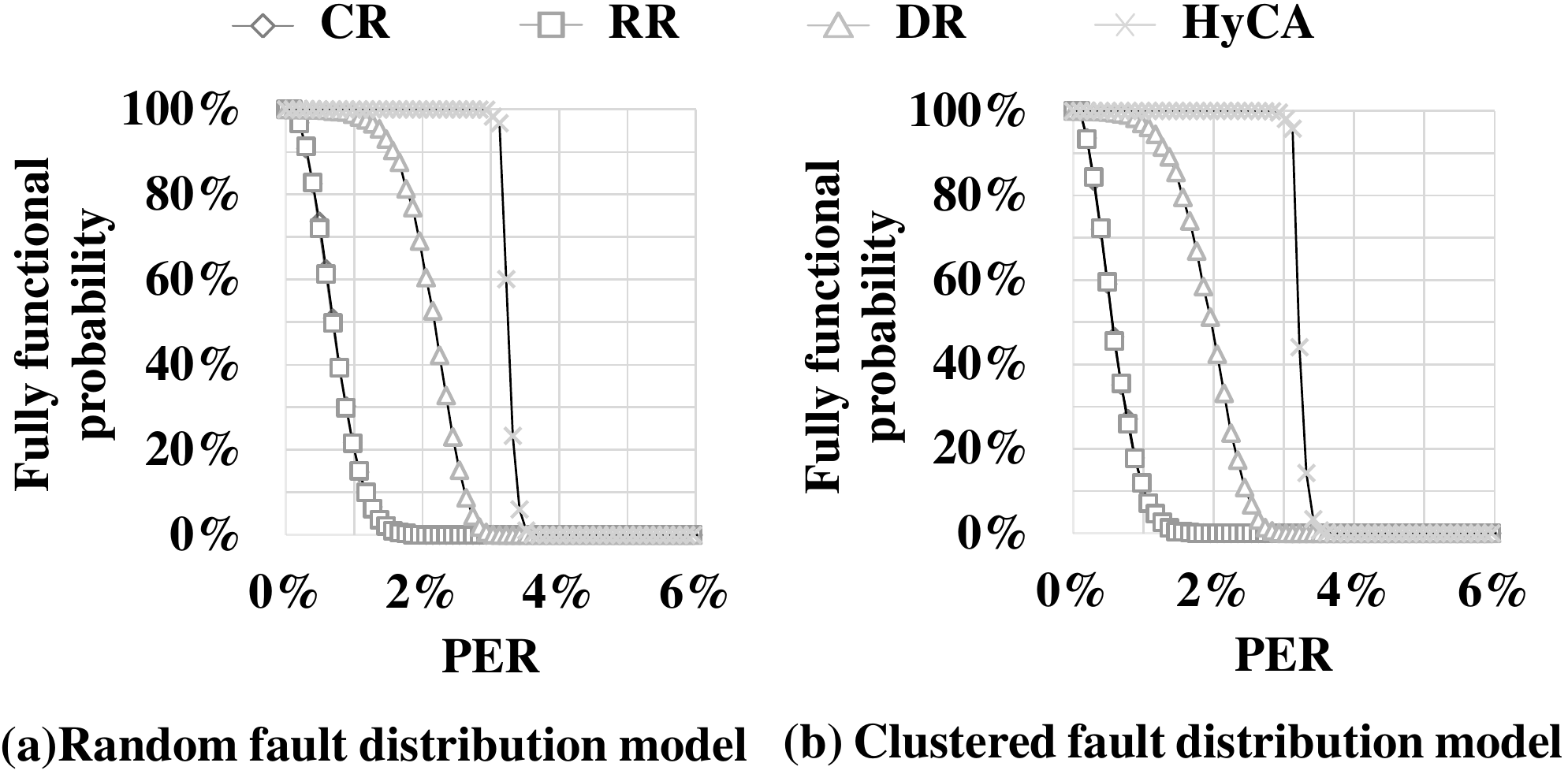}}
    \caption{Fully functional probability of DLAs with different redundancy approaches.}
\label{fig:survival}
\vspace{-1.5em}
\end{figure}

To evaluate the reliability of the DLAs, we propose two metrics that can be applied for different applications. One of them is the fully functional probability and it shows the probability that the DLA can be fully functional without any performance penalty. It is preferred by the mission-critical applications that do not allow any performance degradation nor model modification because any system modification may require expensive and lengthy safety evaluation and certification. The experiment is shown in \autoref{fig:survival}. It shows that HyCA outperforms the three classical redundancy approaches and the advantage gets enlarged under the clustered fault distribution. The main reason is that each redundant PE in RR, CR and DR can only be utilized to replace a single faulty PE in a row, a column, and a row-column pair respectively. When multiple faults occur in the same protected region, these redundancy approaches fail to recover the faulty 2-D computing array and the design will not be fully functional. Unlike these classical redundancy approaches, HyCA allows arbitrary faulty distribution and can perfectly repair the computing array as long as the number of faulty PEs in the 2-D computing array does not exceed the DPPU size. Thereby, the fully function probability of HyCA is not sensitive to the fault distribution models. As DPPU size is set to be 32 and the 2-D computing array size is $32 \times 32$ in this example, the fully functional probability drops to 0 immediately when the number of fault PEs exceeds 32 at 3.13\% PER. As the PEs in the DPPU can also be faulty, the fully functional probability of HyCA starts to drop when the number of faulty PE is close to 32 and the PER is slightly lower than 3.13\%.

The other metric is the normalized remaining computing power and it refers to the percentage of the remaining computing array size over the original 2-D computing array size. This metric is particularly important for the non-critical applications that do not require fully functional accelerators and allow the accelerators to be degraded, because the remaining computing array size determines the theoretical computing power and affects the performance of the deployed neural network models directly. In this work, we apply the acceleration degradation strategy mentioned in the end of Section \ref{sec:dataflow} and discard the faulty PEs in the granularity of a column when the redundant PEs are insufficient to mitigate all the faulty PEs. Although more aggressive degradation approaches are possible to achieve larger computing power, this approach is applied for more efficient model compilation, hardware implementation and memory accesses. 

\begin{figure}
\setlength{\abovecaptionskip}{-10pt}
\setlength{\belowcaptionskip}{-2pt}
	\center{\includegraphics[width=0.99\linewidth]{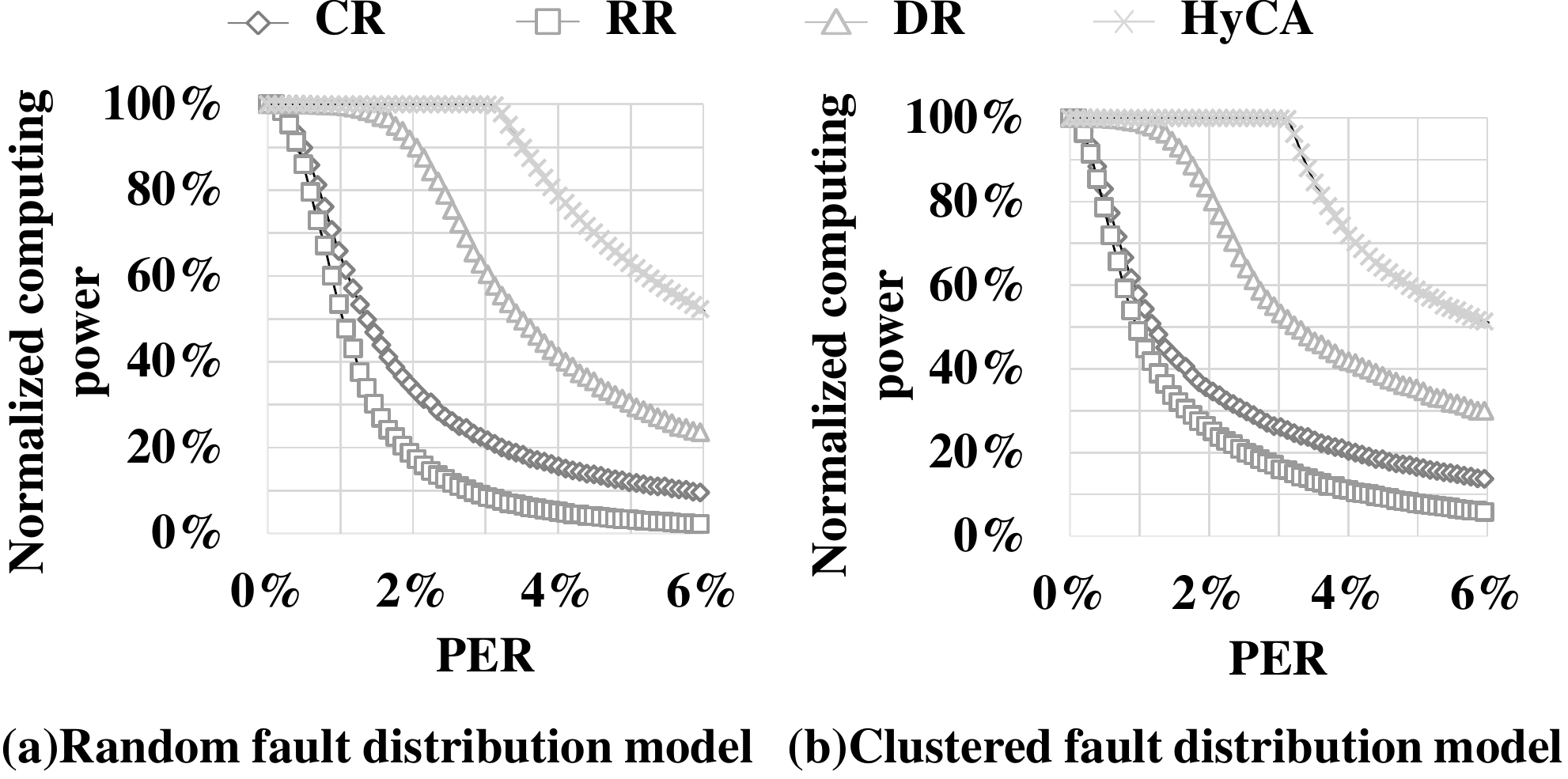}}
    \caption{Normalized computing power of DLAs with different redundancy approaches.}
\label{fig:available}
\vspace{-1em}
\end{figure}


\autoref{fig:available} reveals the computing power comparison of the different redundancy approaches. It can be observed that HyCA shows significantly higher computing power under all the different PER setups and the advantage also enlarges with the increase of the PER. This is mainly brought by the fault recovery flexibility of the HyCA that allows the DPPU to select the most critical faulty PEs to repair when the redundant faulty PEs are insufficient. Note that the most critical faulty PEs refer to the ones that can maximize the remaining 2-D computing array. By optimizing the faulty PE mitigation order, the remaining computing array can be larger. In contrast, each redundant PE can only repair a limited subset of the faulty PEs for the RR, CR and DR. There is little space left to optimize the faulty PE mitigation order. Thereby, the remaining computing power of RR, CR, and DR is much lower. As we choose to discard the faulty PEs that are failed to be repaired in the granularity of a column, RR cannot effectively shift the faulty PEs to a different column and has to discard the column whenever there are more than one faulty PEs. As a result, RR shows the lowest computing power even when the number of redundant PEs is the same with the other redundancy approaches.

\begin{figure}
\setlength{\abovecaptionskip}{-10pt}
\setlength{\belowcaptionskip}{-20pt}
	\center{\includegraphics[width=1\linewidth]{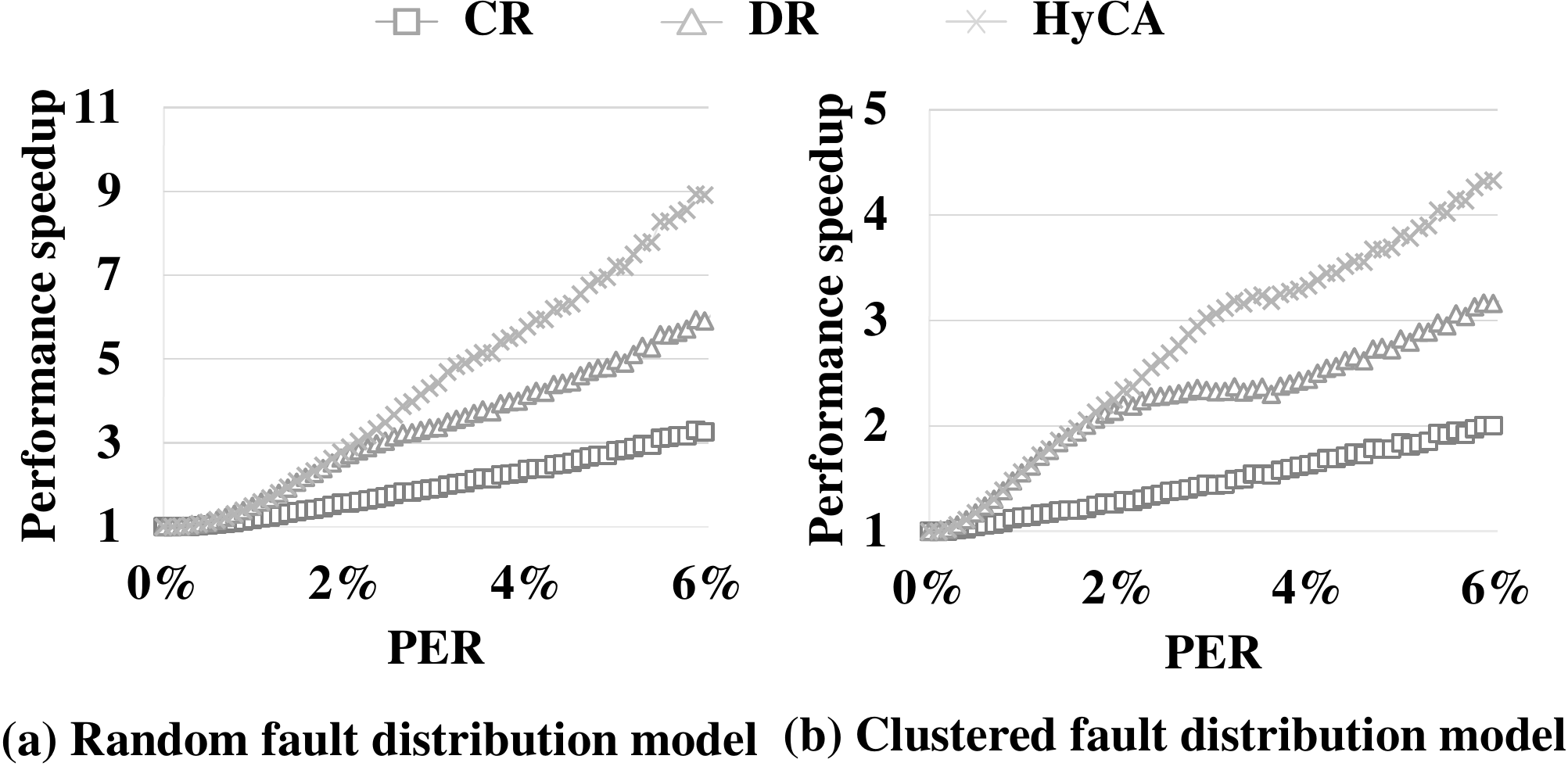}}
    \caption{Normalized performance of DLAs with different redundancy approaches}
\label{fig:ratio}
\vspace{-1em}
\end{figure}

\begin{figure}
\setlength{\abovecaptionskip}{-2pt}
\setlength{\belowcaptionskip}{-10pt}
	\center{\includegraphics[width=0.8\linewidth]{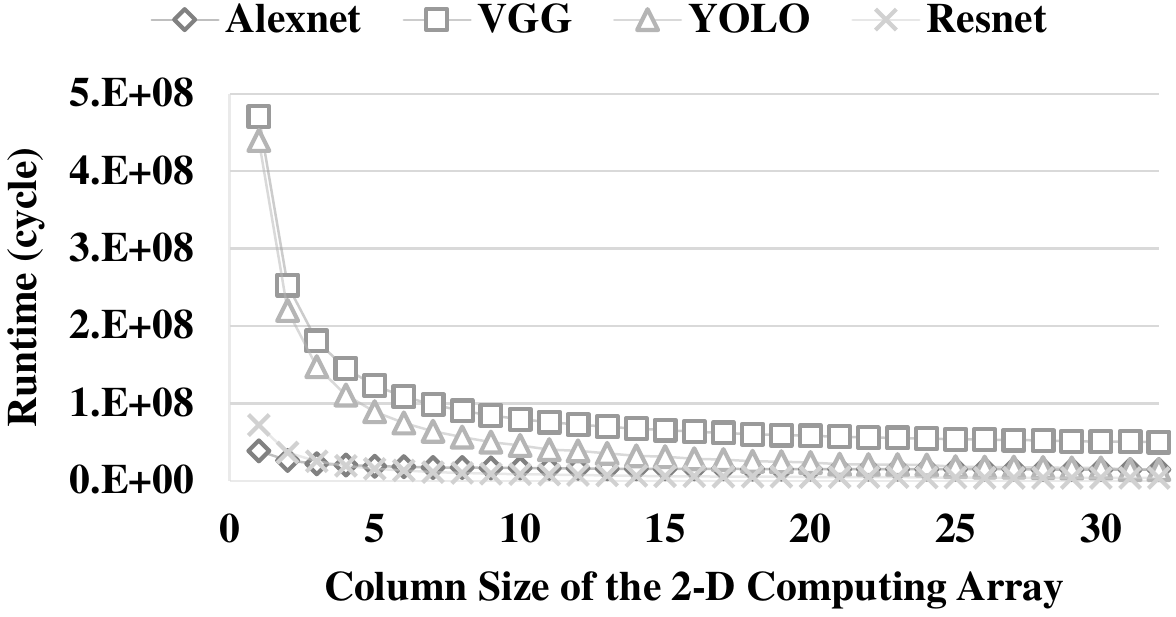}}
    \caption{Neural network runtime of the DLAs with different computing array sizes. Note that the row size of the computing arrays is fixed to be 32.}
\label{fig:model-performance}
\vspace{-1em}
\end{figure}

\begin{figure*}
\setlength{\abovecaptionskip}{-10pt}
\setlength{\belowcaptionskip}{0pt}
	\center{\includegraphics[width=1\linewidth]{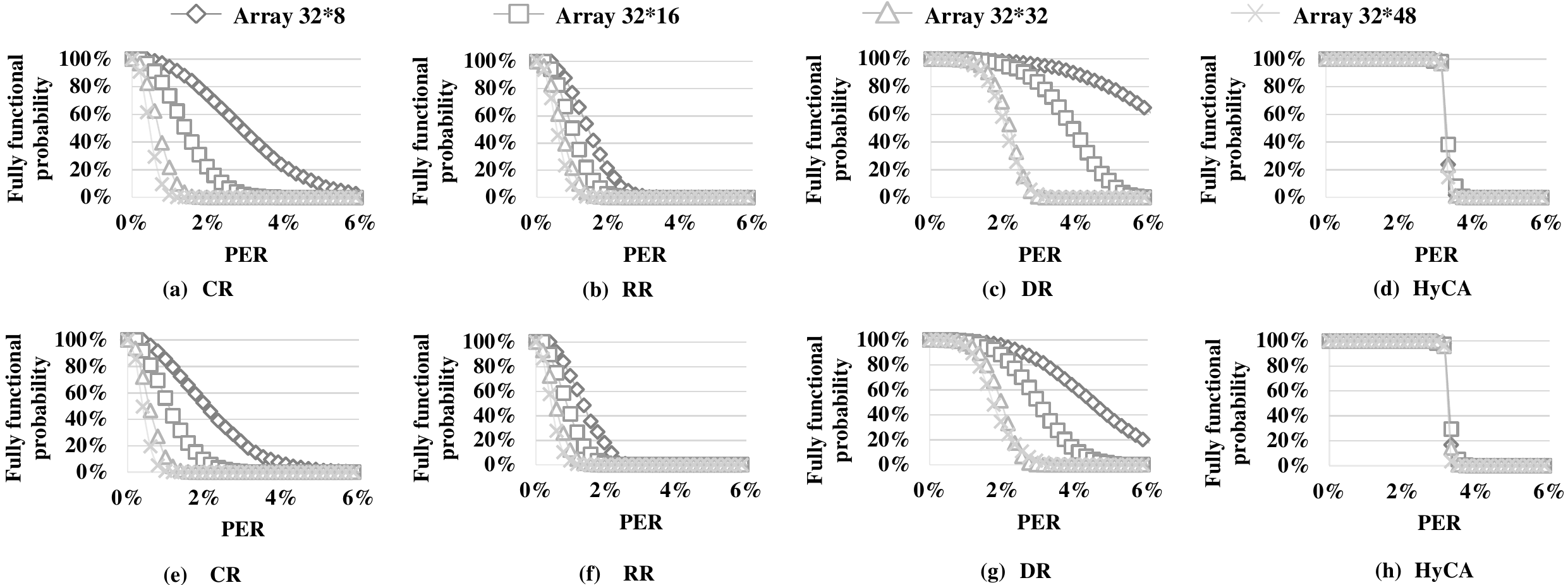}}
    \caption{Fully functional probability of the DLAs with different computing array sizes when they are protected with RR, CR, DR and HyCA respectively. Figure(a-d) are evaluated under the random fault distribution while Figure (e-h) are evaluated under the clustered fault distribution.}
\label{fig:scalability}
\vspace{-1em}
\end{figure*}

\subsection{Performance Comparison}
In order to evaluate the performance of the DLAs protected with the different redundancy approaches, we have the neural network benchmark deployed on the DLAs with Scale-Sim. The performance is normalized to that of the DLA protected with RR and the experiment result is shown in \autoref{fig:ratio}. It can be found that HyCA achieves much higher performance than the other redundancy approaches especially under relatively higher PER, which is roughly consistent with the experiment in \autoref{fig:available} though the neural networks also affect the performance. Particularly, the performance speedup goes up to 9X when the PER is around 6\% under the random fault distribution. The underlying reason for the higher performance speedup at higher PER is that higher PER indicates more faulty PEs in the 2-D computing array and leaves larger optimization space for the HyCA. 

Another observation is that the performance gap between HyCA and the other redundancy approaches is much smaller than the computing power gap. For instance, the computing power of HyCA is around 25X higher than RR when PER is 6\% under the random fault injection while the performance speedup is only 9X. This is mainly attributed to the fact that the neural network runtime varies dramatically under different computing array sizes as shown in \autoref{fig:model-performance}. When the remaining computing array size is large at lower PER, the runtime decreases much slower with the increase of the computing array size. In addition, some of the neural networks like VGG include some full connection layers that fail to make best use of the computing array. In fact, only a single column of PEs is used for the full connection operations given the output stationary dataflow and the larger remaining computing array in HyCA is underutilized, which also undermines the performance speedup.

\subsection{Redundancy Design Scalability Analysis}
As different applications may have distinct requirements of reliability and may also work under different fault environments, a scalable redundancy design can greatly alleviate the reliability design problems. In this subsection, we will investigate and compare the scalability of the different redundancy approaches under different computing array sizes. As the fully functional probability and the computing power is roughly positively related, we only use the fully function probability as the metric for the scalability evaluation to save the space. The number of redundant PEs in RR and CR is consistent with the corresponding computing array row size and column size respectively. Although the number of redundant PEs in DR is also equal to the diagonal size of the computing array, it cannot be directly applied to a non-square computing array. In this experiment, we divide the non-square computing array into multiple square computing arrays and apply the diagonal redundancy approach to each sub computing array independently. The number of redundant PEs in HyCA is set to be $Col$ to ensure a fair comparison where $Col$ refers to the column size of the computing arrays. 

The experiment result is shown in \autoref{fig:scalability}. It can be observed that the fully functional probability of RR, CR and DR under different PER setups vary dramatically when the computing array size changes. CR and DR have the same amount of redundant PEs on the different 2-D computing array sizes. Basically, the redundancy intensity, i.e. the average redundancy per PE in the 2-D computing array, vary dramatically across the different computing arrays. Thus, the fully functional probability curves are different accordingly. The number of redundant PEs for RR scales with the three specific computing arrays, but the fully functional probability curves are closer to each other but remains different due to the fault distribution variations. In general, the classical redundancy approaches do not scale well and the sensitivity to the fault distribution further aggravates the scalability problem. In contrast, the proposed HyCA exhibits much better scalability and shows consistent fault-tolerance capability under different computing array sizes and fault distribution models.

\begin{figure}
\setlength{\abovecaptionskip}{-10pt}
\setlength{\belowcaptionskip}{0pt}
	\center{\includegraphics[width=1\linewidth]{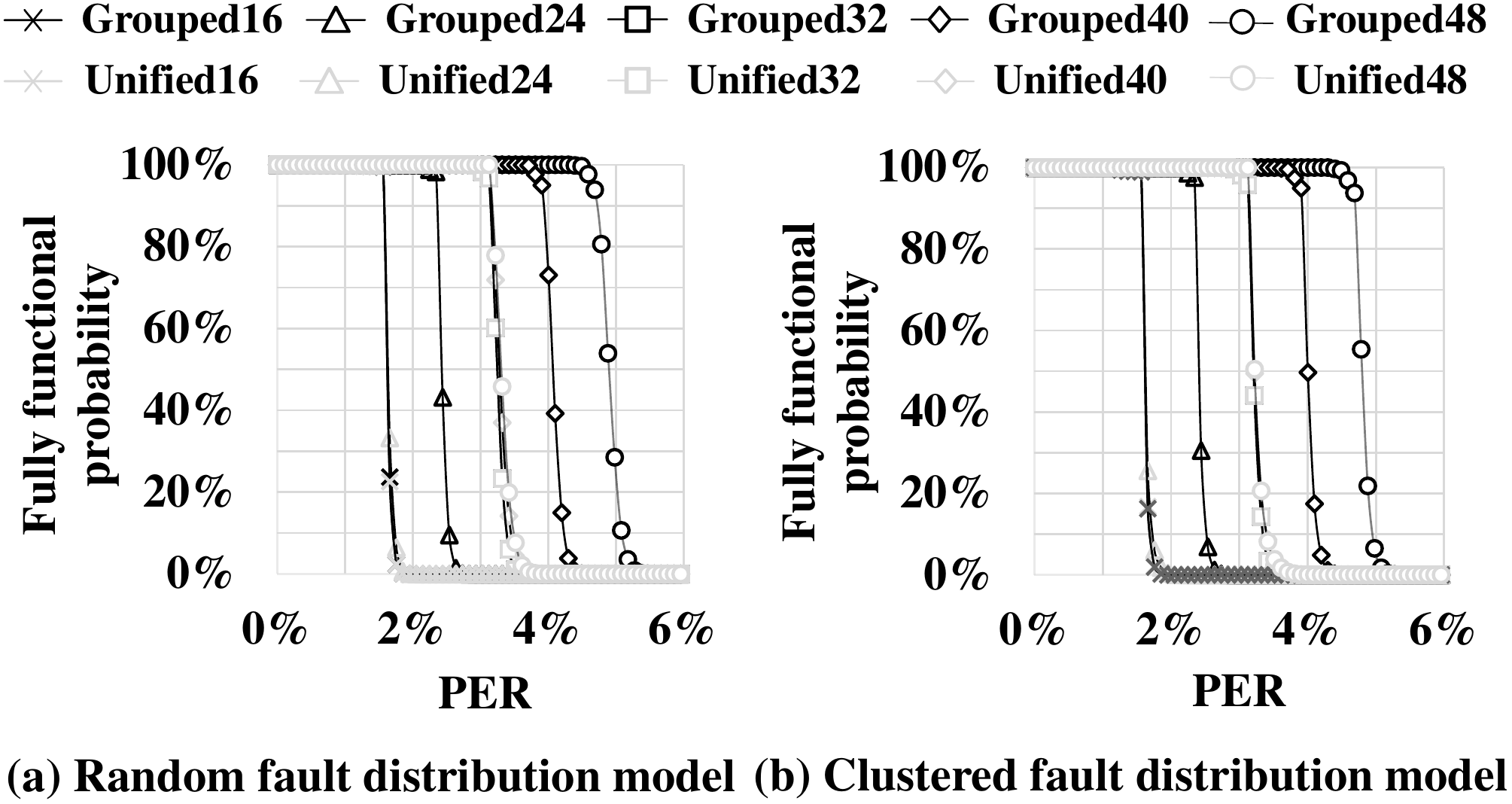}}
    \caption{Fully functional probability of the DLAs configured with different DPPU sizes. The DPPUs with both the Unified structure and the Grouped structure are evaluated and compared.}
\label{fig:DPPU}
\vspace{-1.5em}
\end{figure}

We also evaluated the scalability of the two different DPPU implementations, i.e. the Unified DPPU and the Grouped DPPU under different DPPU sizes. We scale the DPPU sizes from 16 to 48 and fix the computing array size to be $32 \times 32$. The experiment result is shown in \autoref{fig:DPPU}. It can be observed that the Grouped DPPU scales strictly with the DPPU sizes. The Unified DPPU scales when the DPPU size is set to be 16 and 32, but it does not scale when the DPPU size is set to be 24, 40, and 48. The main reason is that the Unified DPPU needs to read from input and weight register files in which the data are aligned with the column size of the computing array. More specifically, the input features and weights are aligned to 32, i.e. the column size. When the DPPU size is larger than 32 and it cannot be perfectly split by 32, the Unified DPPU cannot be fully utilized due to the lack of the sufficient input data. When the DPPU size is smaller than 32 and 32 data cannot be divided perfectly for the Unified DPPU processing, the Unified DPPU also suffers underutilization and leads to unsatisfactory scalability in this occasion. Different from the Unified DPPU, the Grouped DPPU can be utilized with smaller granularity and is able to make full use of the aligned data from both the input and weight register files. As the hardware overhead of DPPU is mainly caused by the redundant PEs according to the experiment in Section \ref{sec:chip-area}, the hardware overhead of both DPPU implementations scales with the DPPU size accordingly.

\subsection{Fault Detection Analysis}
With the proposed fault detection approach, PE faults can be detected at runtime. Since the proposed fault detection module essentially scans all the PEs in the 2-D computing array sequentially, we mainly evaluate the fault detection scanning time under different computing array sizes, and compare the fault detection time to the corresponding neural network processing time. Basically, we want to determine if a runtime persistent fault can be detected before a neural network layer is computed. And we take the percentage of the neural network layers that can be detected during the layer processing as a metric to evaluate the fault detection capability. The experiment result is shown in \autoref{detection}. It can be observed that all the faults in the 2-D computing array can be detected during the execution of each neural network layer when the 2-D computing array size is smaller than or equal to $64 \times 64$. When the 2-D computing array size reaches $128 \times 128$, the processing time of some small neural network layers can finish the processing before the fault detection module scan the entire 2-D computing array. In this case, we may have to add more DPPU groups for the fault detection. 

The fault detection module mainly includes a check list buffer (CLB) and some control logic. The CLB is $Col \times W \times 4$ bytes and dominates the chip area, but it is only $Row/(2\times W)$ (i.e. 1/4 when $Row = 32$ and $W = 4$) of the input register file. Thus, the CLB overhead is much smaller than the input register file let alone the redundant PEs. Thereby, the chip area of the fault detection module is negligible.

\begin{table}[]
\setlength{\abovedisplayskip}{3pt}
\setlength{\belowdisplayskip}{3pt}
\centering
\setlength{\tabcolsep}{4mm}
\caption{The proportion of the neural network layers of which the execution time can fully cover the fault detection of the entire 2-D computing array.}
\begin{tabular}{ccccc}
\hline
\hline
 Array Size       & 16$\times$16  & 32$\times$32  & 64$\times$64  & 128$\times$128 \\     \hline
Alexnet & 8/8           & 8/8           & 8/8           & 4/8     \\            
VGG     & 16/16         & 16/16         & 16/16         & 16/16   \\            
YOLO    & 22/22         & 22/22         & 22/22         & 15/22   \\            
Resnet  & 21/21         & 21/21         & 21/21         & 5/21    \\            \hline
\end{tabular}
\label{detection}
\vspace{-1em}
\end{table}

\section{conclusion} \label{sec:conclusion}
The reliability of DLAs is of vital importance to the mission-critical AI applications. Prior redundancy design approaches for the regular computing array such as RR and CR greatly reduce the hardware overhead compared to the classical TMR approaches, but they are rather sensitive to the fault distribution and fail to work especially when the faults are unevenly distributed. To address this problem, we propose a hybrid computing architecture (HyCA) and have a DPPU to recompute all the operations mapped to the faulty PEs in the 2-D computing array. When the number of faulty PEs in the 2-D computing array is less than the DPPU size, HyCA can fully recover the 2-D computing array despite the fault distribution. Even when the fault error rate further increases, DPPU can still be used to repair the most critical PEs first to ensure a large available computing array and minimize the performance penalty. According to our experiments, HyCA outperforms prior redundancy approaches in terms of both the fully functional probability and the computing power under different fault distribution models. In addition, HyCA can also be reused for the fault detection at runtime and the experiment result shows that the entire 2-D computing array can be scanned and detected before a neural network layer completes its execution in most cases.

\section{Acknowledgement}\label{acknowledgement}

This work is supported in part by National Key
Research and Development Program of China under Grant No.2020YFB1600201, National Natural Science Foundation of China under Grant No.62174162, No.61902375 and No.61834006. The corresponding author is Dawen Xu.
\bibliographystyle{IEEEtran}
\bibliography{refs}

\begin{thebibliography}{10}
\providecommand{\url}[1]{#1}
\csname url@samestyle\endcsname
\providecommand{\newblock}{\relax}
\providecommand{\bibinfo}[2]{#2}
\providecommand{\BIBentrySTDinterwordspacing}{\spaceskip=0pt\relax}
\providecommand{\BIBentryALTinterwordstretchfactor}{4}
\providecommand{\BIBentryALTinterwordspacing}{\spaceskip=\fontdimen2\font plus
\BIBentryALTinterwordstretchfactor\fontdimen3\font minus
  \fontdimen4\font\relax}
\providecommand{\BIBforeignlanguage}[2]{{%
\expandafter\ifx\csname l@#1\endcsname\relax
\typeout{** WARNING: IEEEtran.bst: No hyphenation pattern has been}%
\typeout{** loaded for the language `#1'. Using the pattern for}%
\typeout{** the default language instead.}%
\else
\language=\csname l@#1\endcsname
\fi
#2}}
\providecommand{\BIBdecl}{\relax}
\BIBdecl

\bibitem{fink2019deep}
M.~Fink, Y.~Liu, A.~Engstle, and S.-A. Schneider, ``{{Deep Learning-Based
  Multi-scale Multi-object Detection and Classification for Autonomous
  Driving}},'' in \emph{Fahrerassistenzsysteme 2018}.\hskip 1em plus 0.5em
  minus 0.4em\relax Springer, 2019, pp. 233--242.

\bibitem{tzelepi2017human}
M.~Tzelepi and A.~Tefas, ``{Human Crowd Detection for Drone Flight Safety Using
  Convolutional Neural Networks},'' in \emph{2017 25th European Signal
  Processing Conference (EUSIPCO)}.\hskip 1em plus 0.5em minus 0.4em\relax
  IEEE, 2017, pp. 743--747.

\bibitem{esteva2019guide}
A.~Esteva, A.~Robicquet, B.~Ramsundar, V.~Kuleshov, M.~DePristo, K.~Chou,
  C.~Cui, G.~Corrado, S.~Thrun, and J.~Dean, ``{A Guide to Deep Learning in
  Healthcare},'' \emph{Nature medicine}, vol.~25, no.~1, pp. 24--29, 2019.

\bibitem{banerjee2019towards}
S.~S. Banerjee, J.~Cyriac, S.~Jha, Z.~T. Kalbarczyk, and R.~K. Iyer, ``{Towards
  a Bayesian Approach for Assessing Fault Tolerance of Deep Neural Networks},''
  in \emph{2019 49th Annual IEEE/IFIP International Conference on Dependable
  Systems and Networks--Supplemental Volume (DSN-S)}.\hskip 1em plus 0.5em
  minus 0.4em\relax IEEE, 2019, pp. 25--26.

\bibitem{jha2019ml}
S.~Jha, S.~Banerjee, T.~Tsai, S.~K. Hari, M.~B. Sullivan, Z.~T. Kalbarczyk,
  S.~W. Keckler, and R.~K. Iyer, ``{ML-Based Fault Injection for Autonomous
  Vehicles: A Case for Bayesian Fault Injection},'' in \emph{2019 49th Annual
  IEEE/IFIP International Conference on Dependable Systems and Networks
  (DSN)}.\hskip 1em plus 0.5em minus 0.4em\relax IEEE, 2019, pp. 112--124.

\bibitem{jenihhin2019challenges}
M.~Jenihhin, M.~S. Reorda, A.~Balakrishnan, and D.~Alexandrescu, ``{Challenges
  of Reliability Assessment and Enhancement in Autonomous Systems},'' in
  \emph{2019 IEEE International Symposium on Defect and Fault Tolerance in VLSI
  and Nanotechnology Systems (DFT)}.\hskip 1em plus 0.5em minus 0.4em\relax
  IEEE, 2019, pp. 1--6.

\bibitem{chen2014dadiannao}
Y.~Chen, T.~Luo, S.~Liu, S.~Zhang, L.~He, J.~Wang, L.~Li, T.~Chen, Z.~Xu,
  N.~Sun \emph{et~al.}, ``{Dadiannao: A Machine-learning Supercomputer},'' in
  \emph{2014 47th Annual IEEE/ACM International Symposium on
  Microarchitecture}.\hskip 1em plus 0.5em minus 0.4em\relax IEEE, 2014, pp.
  609--622.

\bibitem{xu2020hybrid}
D.~Xu, C.~Chu, Q.~Wang, C.~Liu, Y.~Wang, L.~Zhang, H.~Liang, and K.-T. Cheng,
  ``{A Hybrid Computing Architecture for Fault-tolerant Deep Learning
  Accelerators},'' in \emph{2020 IEEE 38th International Conference on Computer
  Design (ICCD)}.\hskip 1em plus 0.5em minus 0.4em\relax IEEE, 2020, pp.
  478--485.

\bibitem{reagen2016minerva}
B.~Reagen, P.~Whatmough, R.~Adolf, S.~Rama, H.~Lee, S.~K. Lee, J.~M.
  Hern{\'a}ndez-Lobato, G.-Y. Wei, and D.~Brooks, ``{Minerva: Enabling
  Low-power, Highly-accurate Deep neural Network Accelerators},'' in \emph{2016
  ACM/IEEE 43rd Annual International Symposium on Computer Architecture
  (ISCA)}.\hskip 1em plus 0.5em minus 0.4em\relax IEEE, 2016, pp. 267--278.

\bibitem{impact2011dixit}
A.~{Dixit} and A.~{Wood}, ``{The Impact of New Technology on Soft Error
  Rates},'' in \emph{2011 International Reliability Physics Symposium}, 2011,
  pp. 5B.4.1--5B.4.7.

\bibitem{mittal2020survey}
S.~Mittal, ``{A Survey on Modeling and Improving Reliability of DNN Algorithms
  and Accelerators},'' \emph{Journal of Systems Architecture}, vol. 104, p.
  101689, 2020.

\bibitem{jouppi2017datacenter}
N.~P. Jouppi, C.~Young, N.~Patil, D.~Patterson, G.~Agrawal, R.~Bajwa, S.~Bates,
  S.~Bhatia, N.~Boden, A.~Borchers \emph{et~al.}, ``{In-datacenter Performance
  Analysis of A Tensor Processing Unit},'' in \emph{Proceedings of the 44th
  Annual International Symposium on Computer Architecture}, 2017, pp. 1--12.

\bibitem{Chen2016Eyeriss}
Y.-H. Chen, J.~Emer, and V.~Sze, ``{Eyeriss: A Spatial Architecture for
  Energy-Efficient Dataflow for Convolutional Neural Networks},'' in \emph{2016
  ACM/IEEE 43rd Annual International Symposium on Computer Architecture
  (ISCA)}.\hskip 1em plus 0.5em minus 0.4em\relax IEEE, 2016, pp. 367--379.

\bibitem{deng2015retraining}
J.~Deng, Y.~Rang, Z.~Du, Y.~Wang, H.~Li, O.~Temam, P.~Ienne, D.~Novo, X.~Li,
  Y.~Chen \emph{et~al.}, ``{Retraining-based timing error mitigation for
  hardware neural networks},'' in \emph{2015 Design, Automation \& Test in
  Europe Conference \& Exhibition (DATE)}.\hskip 1em plus 0.5em minus
  0.4em\relax IEEE, 2015, pp. 593--596.

\bibitem{zhang2019fault}
J.~J. Zhang, K.~Basu, and S.~Garg, ``{Fault-Tolerant Systolic Array Based
  Accelerators for Deep Neural Network Execution},'' \emph{IEEE Design \&
  Test}, vol.~36, no.~5, pp. 44--53, 2019.

\bibitem{xu2019resilient}
D.~Xu, K.~Xing, C.~Liu, Y.~Wang, Y.~Dai, L.~Cheng, H.~Li, and L.~Zhang,
  ``{Resilient Neural Network Training for Accelerators with Computing
  Errors},'' in \emph{2019 IEEE 30th International Conference on
  Application-specific Systems, Architectures and Processors (ASAP)}, vol.
  2160.\hskip 1em plus 0.5em minus 0.4em\relax IEEE, 2019, pp. 99--102.

\bibitem{li2019squeezing}
L.~Li, D.~Xu, K.~Xing, C.~Liu, Y.~Wang, H.~Li, and X.~Li, ``{Squeezing the Last
  {MHz} for CNN Acceleration on {FPGAs}},'' in \emph{2019 IEEE International
  Test Conference in Asia (ITC-Asia)}.\hskip 1em plus 0.5em minus 0.4em\relax
  IEEE, 2019, pp. 151--156.

\bibitem{validation2019Ebert}
C.~{Ebert} and M.~{Weyrich}, ``{Validation of Autonomous Systems},'' \emph{IEEE
  Software}, vol.~36, no.~5, pp. 15--23, 2019.

\bibitem{takanami2012built}
I.~Takanami and T.~Horita, ``{A Built-in Circuit for Self-Repairing
  Mesh-Connected Processor Arrays by Direct Spare Replacement},'' in \emph{2012
  IEEE 18th Pacific Rim International Symposium on Dependable Computing}.\hskip
  1em plus 0.5em minus 0.4em\relax IEEE, 2012, pp. 96--104.

\bibitem{takanami2017built}
I.~{Takanami} and M.~{Fukushi}, ``{A Built-in Circuit for Self-Repairing
  Mesh-Connected Processor Arrays with Spares on Diagonal},'' in \emph{2017
  IEEE 22nd Pacific Rim International Symposium on Dependable Computing
  (PRDC)}, 2017, pp. 110--117.

\bibitem{xu2021reliability}
D.~{Xu}, Z.~{Zhu}, C.~{Liu}, Y.~{Wang}, S.~{Zhao}, L.~{Zhang}, H.~{Liang},
  H.~{Li}, and K.~T. {Cheng}, ``{Reliability Evaluation and Analysis of
  FPGA-Based Neural Network Acceleration System},'' \emph{IEEE Transactions on
  Very Large Scale Integration (VLSI) Systems}, pp. 1--13, 2021.

\bibitem{xu2020persistent}
D.~{Xu}, Z.~{Zhu}, C.~{Liu}, Y.~{Wang}, H.~{Li}, L.~{Zhang}, and K.~{Cheng},
  ``{Persistent Fault Analysis of Neural Networks on FPGA-based Acceleration
  System},'' in \emph{2020 IEEE 31st International Conference on
  Application-specific Systems, Architectures and Processors (ASAP)}, 2020, pp.
  85--92.

\bibitem{ning2020ftt}
W.~Li, X.~Ning, G.~Ge, X.~Chen, Y.~Wang, and H.~Yang, ``Ftt-nas: Discovering
  fault-tolerant neural architecture,'' in \emph{2020 25th Asia and South
  Pacific Design Automation Conference (ASP-DAC)}.\hskip 1em plus 0.5em minus
  0.4em\relax IEEE Press, 2020, p. 211–216.

\bibitem{error2018date}
M.~A. {Hanif}, R.~{Hafiz}, and M.~{Shafique}, ``{Error Resilience Analysis for
  Systematically Employing Approximate Computing in Convolutional Neural
  Networks},'' in \emph{2018 Design, Automation Test in Europe Conference
  Exhibition (DATE)}, 2018, pp. 913--916.

\bibitem{energy2018kim}
S.~{Kim}, P.~{Howe}, T.~{Moreau}, A.~{Alaghi}, L.~{Ceze}, and V.~S. {Sathe},
  ``{Energy-Efficient Neural Network Acceleration in the Presence of Bit-Level
  Memory Errors},'' \emph{IEEE Transactions on Circuits and Systems I: Regular
  Papers}, vol.~65, no.~12, pp. 4285--4298, 2018.

\bibitem{axtrain2018he}
X.~{He}, W.~{Lu}, G.~{Yan}, and X.~{Zhang}, ``{Joint Design of Training and
  Hardware Towards Efficient and Accuracy-Scalable Neural Network Inference},''
  \emph{IEEE Journal on Emerging and Selected Topics in Circuits and Systems},
  vol.~8, no.~4, pp. 810--821, 2018.

\bibitem{wang2017resilience}
Y.~Wang, J.~Deng, Y.~Fang, H.~Li, and X.~Li, ``{Resilience-aware Frequency
  Tuning for Neural-network-based Approximate Computing Chips},'' \emph{IEEE
  Transactions on Very Large Scale Integration (VLSI) Systems}, vol.~25,
  no.~10, pp. 2736--2748, 2017.

\bibitem{analyzing2018vts}
J.~J. {Zhang}, T.~{Gu}, K.~{Basu}, and S.~{Garg}, ``{Analyzing and Mitigating
  the Impact of Permanent Faults on a Systolic Array based Neural Network
  Accelerator},'' in \emph{2018 IEEE 36th VLSI Test Symposium (VTS)}, 2018, pp.
  1--6.

\bibitem{abdullah2020salvagednn}
M.~Abdullah~Hanif and M.~Shafique, ``{Salvagednn: Salvaging Deep Neural Network
  Accelerators with Permanent Faults Through Saliency-driven Fault-aware
  Mapping},'' \emph{Philosophical Transactions of the Royal Society A}, vol.
  378, no. 2164, p. 20190164, 2020.

\bibitem{horita2000fault}
T.~Horita and I.~Takanami, ``{Fault-tolerant Processor Arrays Based on the 1
  1/2-track Switches with Flexible Spare Distributions},'' \emph{IEEE
  Transactions on Computers}, vol.~49, no.~6, pp. 542--552, 2000.

\bibitem{stapper1983integrated}
C.~H. Stapper, F.~M. Armstrong, and K.~Saji, ``{Integrated Circuit Yield
  Statistics},'' \emph{Proceedings of the IEEE}, vol.~71, no.~4, pp. 453--470,
  1983.

\bibitem{ozen2019sanity}
E.~Ozen and A.~Orailoglu, ``{Sanity-Check: Boosting the reliability of
  safety-critical deep neural network applications},'' in \emph{2019 IEEE 28th
  Asian Test Symposium (ATS)}.\hskip 1em plus 0.5em minus 0.4em\relax IEEE,
  2019, pp. 7--75.

\bibitem{zhao2020algorithm}
K.~Zhao, S.~Di, S.~Li, X.~Liang, Y.~Zhai, J.~Chen, K.~Ouyang, F.~Cappello, and
  Z.~Chen, ``{Algorithm-Based Fault Tolerance for Convolutional Neural
  Networks},'' \emph{arXiv preprint arXiv:2003.12203}, 2020.

\bibitem{zhang2020sorting}
Y.~Zhang, S.~Lin, R.~Wang, Y.~Wang, Y.~Wang, W.~Qian, and R.~Huang, ``{When
  Sorting Network Meets Parallel Bitstreams: A Fault-tolerant Parallel Ternary
  Neural Network Accelerator Based on Stochastic Computing},'' in \emph{2020
  Design, Automation \& Test in Europe Conference \& Exhibition (DATE)}.\hskip
  1em plus 0.5em minus 0.4em\relax IEEE, 2020, pp. 1287--1290.

\bibitem{li2020soft}
W.~Li, G.~Ge, K.~Guo, X.~Chen, Q.~Wei, Z.~Gao, Y.~Wang, and H.~Yang, ``{Soft
  Error Mitigation for Deep Convolution Neural Network on FPGA Accelerators},''
  in \emph{2020 2nd IEEE International Conference on Artificial Intelligence
  Circuits and Systems (AICAS)}.\hskip 1em plus 0.5em minus 0.4em\relax IEEE,
  2020, pp. 1--5.

\bibitem{mahdiani2012relaxed}
H.~R. Mahdiani, S.~M. Fakhraie, and C.~Lucas, ``{Relaxed Fault-tolerant
  Hardware Implementation of Neural Networks in the Presence of Multiple
  Transient Errors},'' \emph{IEEE transactions on neural networks and learning
  systems}, vol.~23, no.~8, pp. 1215--1228, 2012.

\bibitem{neggaz2018reliability}
M.~A. Neggaz, I.~Alouani, P.~R. Lorenzo, and S.~Niar, ``{A Reliability Study on
  CNNs for Critical Embedded Systems},'' in \emph{2018 IEEE 36th International
  Conference on Computer Design (ICCD)}.\hskip 1em plus 0.5em minus 0.4em\relax
  IEEE, 2018, pp. 476--479.

\bibitem{ares2018dac}
B.~{Reagen}, U.~{Gupta}, L.~{Pentecost}, P.~{Whatmough}, S.~K. {Lee},
  N.~{Mulholland}, D.~{Brooks}, and G.~{Wei}, ``{Ares: A framework for
  Quantifying the Resilience of Deep Neural Networks},'' in \emph{2018 55th
  ACM/ESDA/IEEE Design Automation Conference (DAC)}, 2018, pp. 1--6.

\bibitem{deng2009imagenet}
J.~Deng, W.~Dong, R.~Socher, L.-J. Li, K.~Li, and L.~Fei-Fei, ``{Imagenet: A
  Large-scale Hierarchical Image Database},'' in \emph{2009 IEEE conference on
  computer vision and pattern recognition}.\hskip 1em plus 0.5em minus
  0.4em\relax IEEE, 2009, pp. 248--255.

\bibitem{Energy2003Aneesh}
{Aneesh Aggarwal} and M.~{Franklin}, ``{Energy Efficient Asymmetrically Ported
  Register Files},'' in \emph{Proceedings 21st International Conference on
  Computer Design}, 2003, pp. 2--7.

\bibitem{A2012Chang}
P.~{Chang}, T.~{Lin}, J.~{Wang}, and Y.~{Yu}, ``{A 4R/2W Register File Design
  for UDVS Microprocessors in 65-nm CMOS},'' \emph{IEEE Transactions on
  Circuits and Systems II: Express Briefs}, vol.~59, no.~12, pp. 908--912,
  2012.

\bibitem{meyer1989modeling}
F.~J. Meyer and D.~K. Pradhan, ``{Modeling Defect Spatial Distribution},''
  \emph{IEEE Transactions on Computers}, vol.~38, no.~4, pp. 538--546, 1989.

\bibitem{liu2011resilient}
C.~Liu, L.~Zhang, Y.~Han, and X.~Li, ``{A resilient on-chip router design
  through data path salvaging},'' in \emph{16th Asia and South Pacific Design
  Automation Conference (ASP-DAC 2011)}.\hskip 1em plus 0.5em minus 0.4em\relax
  IEEE, 2011, pp. 437--442.

\bibitem{li2008understanding}
M.-L. Li, P.~Ramachandran, S.~K. Sahoo, S.~V. Adve, V.~S. Adve, and Y.~Zhou,
  ``{Understanding the propagation of hard errors to software and implications
  for resilient system design},'' \emph{ACM Sigplan Notices}, vol.~43, no.~3,
  pp. 265--276, 2008.

\bibitem{zhang2018analyzing}
J.~J. Zhang, T.~Gu, K.~Basu, and S.~Garg, ``{Analyzing and Mitigating the
  Impact of Permanent Faults on a Systolic Array Based Neural Network
  Accelerator},'' in \emph{2018 IEEE 36th VLSI Test Symposium (VTS)}.\hskip 1em
  plus 0.5em minus 0.4em\relax IEEE, 2018, pp. 1--6.

\bibitem{qian2016optimal}
J.~Qian, Z.~Zhou, T.~Gu, L.~Zhao, and L.~Chang, ``{Optimal Reconfiguration of
  High-performance VLSI Subarrays with Network Flow},'' \emph{IEEE Transactions
  on Parallel and Distributed Systems}, vol.~27, no.~12, pp. 3575--3587, 2016.

\bibitem{samajdar2018scale}
A.~Samajdar, Y.~Zhu, P.~Whatmough, M.~Mattina, and T.~Krishna, ``{Scale-sim:
  Systolic CNN Accelerator Simulator},'' \emph{arXiv preprint
  arXiv:1811.02883}, 2018.

\end{thebibliography}

%

\begin{IEEEbiography}[{\includegraphics[width=1in,height=1.25in,clip,keepaspectratio]{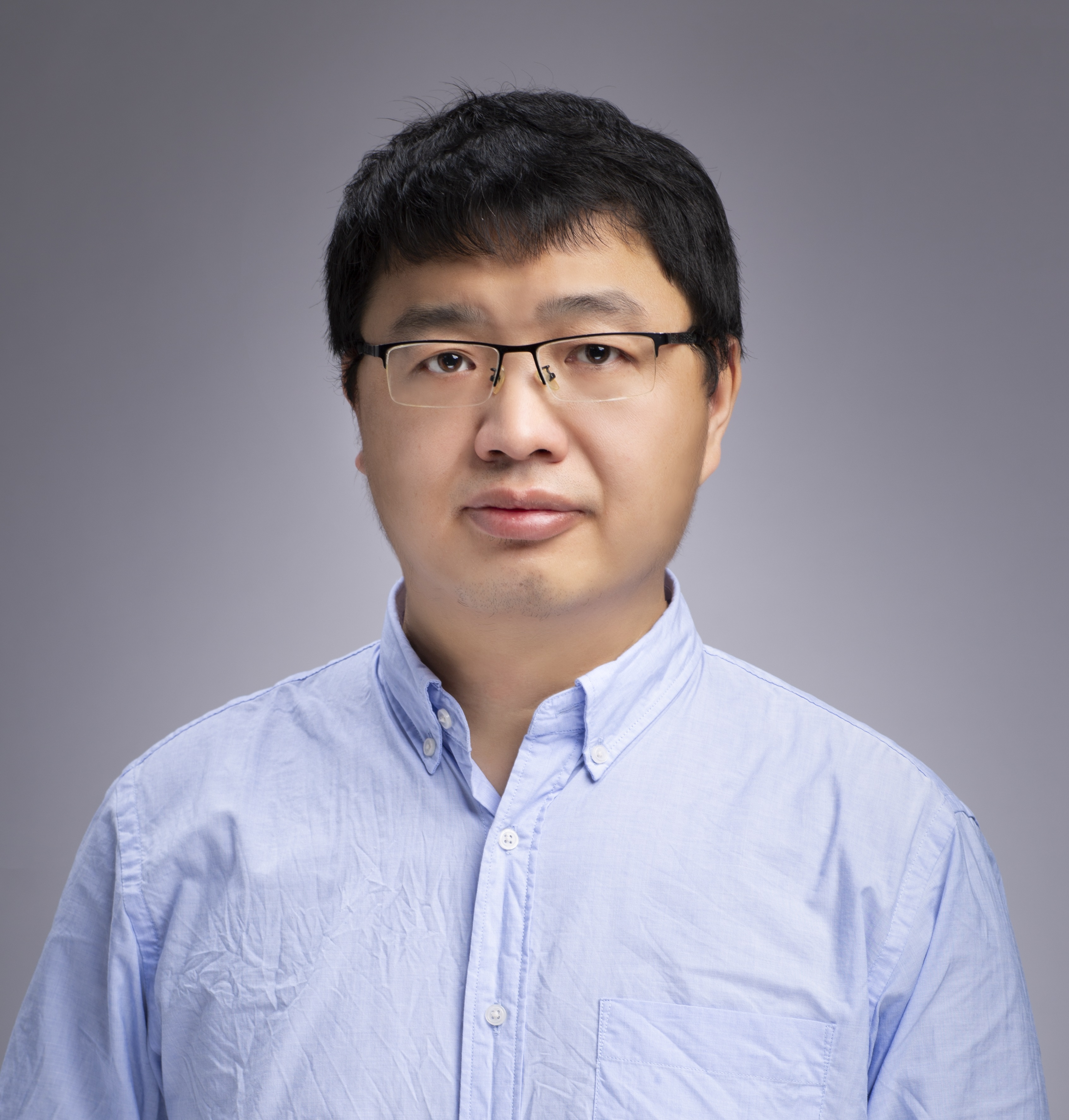}}]{Cheng Liu}
 	received the B.Eng. and M.Eng. degrees from the Harbin Institute of Technology, Harbin, China in 2007 and 2009 respectively, and the Ph.D. degree from the University of Hong Kong in 2016. Currently, he is an associate professor in Institute of Computing Technology (ICT), Chinese Academy of Sciences (CAS). His research interests include FPGA-based reconfigurable computing, fault-tolerant computing, and custom computing.
 \vspace{8mm}
\end{IEEEbiography}

\begin{IEEEbiography}[{\includegraphics[width=1in,height=1.25in,clip,keepaspectratio]{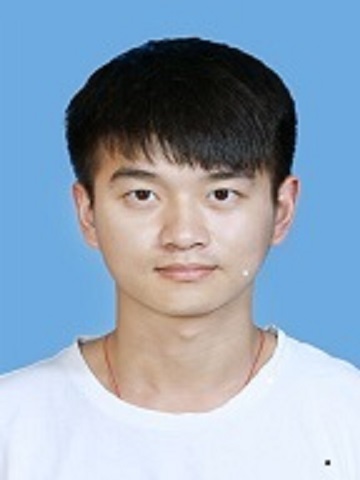}}]{Cheng Chu}
 	received the BS degrees from the Hefei University of Technology, hefei, Anhui, China, in 2018. He is currently a MS of the Hefei University of Technology, hefei, Anhui, China. His research falls primarily in the field of computer architecture, with an emphasis on the fault-tolerant VLSI design, energy-efficient architecture design, and neural network acceleration.
 \vspace{8mm}
\end{IEEEbiography}

\begin{IEEEbiography}[{\includegraphics[width=1in,height=1.25in,clip,keepaspectratio]{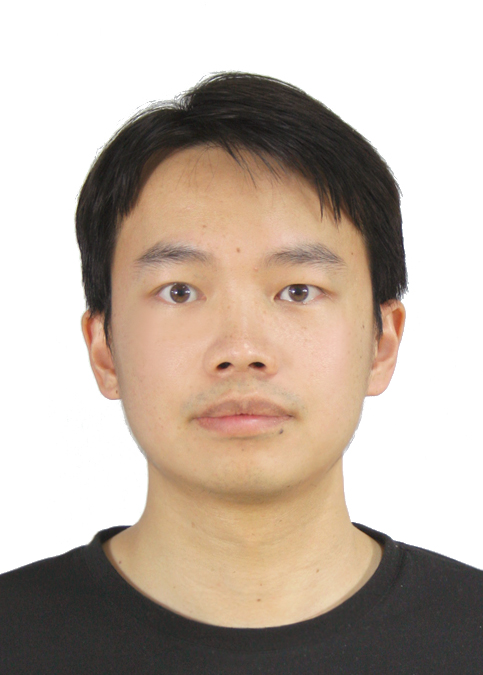}}]{Dawen Xu}
 	received the B.S. degree in computer science from Xi'dian University, Xi'an, China, in 2007, and the M.S. and Ph.D. degrees from the Institute of Computing Technology, Chinese Academy of Sciences, Beijing, China, in 2009 and 2013, respectively. He is currently an associate professor in Hefei University of Technology, Hefei, China. His current research interests include heterogeneous computing, VLSI design and testing, and reliable system.
\end{IEEEbiography}

\begin{IEEEbiography}[{\includegraphics[width=1in,height=1.25in,clip,keepaspectratio]{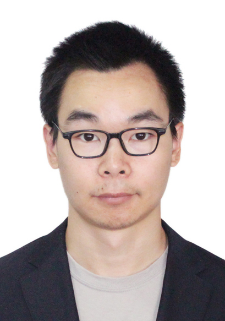}}]{Ying Wang}
	(M’14) received the B.S. and M.S. degrees in electronic engineering from the Harbin Institute of Technology, Harbin, China, in 2007 and 2009, respectively, and the Ph.D. degree in computer science from the Institute of Computing Technology (ICT), Chinese Academy of Sciences (CAS), Beijing, China, in 2014.,He is currently an Associate Professor with ICT, CAS. His current research interests includes computer architecture and VLSI design, specifically memory system, on-chip interconnects, resilient and energy-efficient architecture, and machine learning accelerators.
\vspace{8mm}
\end{IEEEbiography}

\begin{IEEEbiography}[{\includegraphics[width=1in,height=1.25in,clip,keepaspectratio]{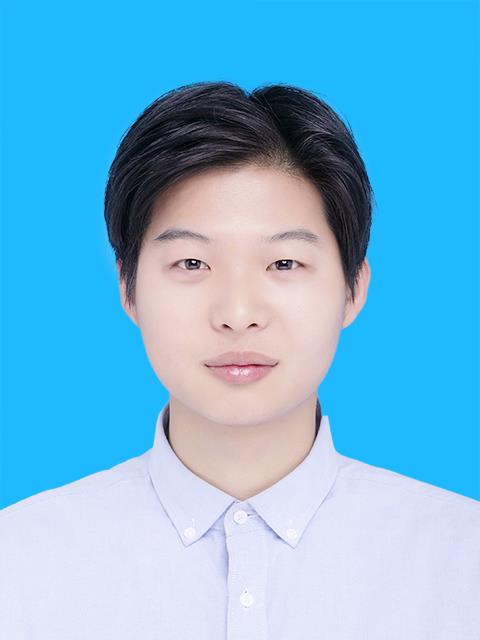}}]{Qianlong Wang}
 	received the BS degrees from the Hefei University of Technology, hefei, Anhui, China, in 2019. He is currently a MS of the Hefei University of Technology, hefei, Anhui, China. His research falls primarily in the field of computer architecture, with an emphasis on the fault-tolerance of CNN accelerator.
\end{IEEEbiography}


\begin{IEEEbiography}[{\includegraphics[width=1in,height=1.25in,clip,keepaspectratio]{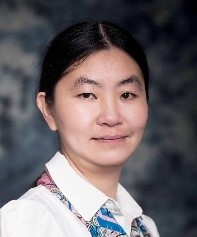}}]{Huawei Li}
    (M’00–SM’09) received the B.S. degree in computer science from Xiangtan University, Xiangtan, China, in 1996, and the M.S. and Ph.D. degrees from the Institute of Computing Technology (ICT), Chinese Academy of Sciences (CAS), Beijing, China, in 1999 and 2001, respectively. She has been a Professor with ICT, CAS since 2008. She visited the University of California at Santa Barbara, Santa Barbara, CA, USA, from 2009 to 2010. She has published over 180 technical papers, and holds 27 Chinese Patents. Her current research interests include testing of VLSI/SOC circuits, design verification, design for reliability, fault-tolerance, and approximate computing. 
    
\end{IEEEbiography}

\begin{IEEEbiography}[{\includegraphics[width=1in,height=1.25in,clip,keepaspectratio]{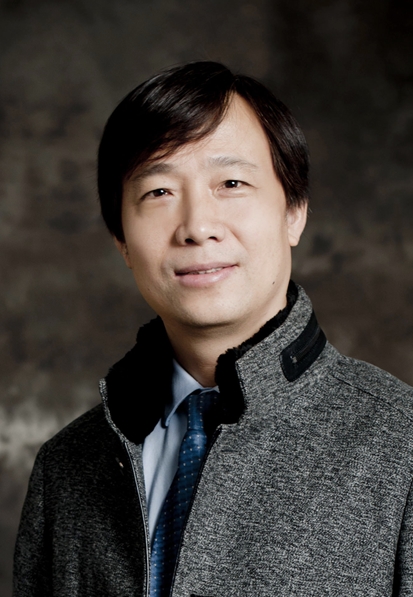}}]{Xiaowei Li}
    Xiaowei Li (SM’04) received the B.Eng. and M.Eng. degrees in computer science from the Hefei University of Technology, Hefei, China, in 1985 and 1988, respectively, and the Ph.D. degree in computer science from the Institute of Computing Technology (ICT), Chinese Academy of Sciences (CAS), Beijing, China, in 1991. He was an Associate Professor with the Department of Computer Science and Technology, Peking University, Beijing, from 1991 to 2000. In 2000, he joined ICT, CAS, as a Professor, where he is currently the Deputy Director of the State Key Laboratory of Computer Architecture. He has coauthored over 280 papers in journals and international conferences, and he holds 60 patents and 30 software copyrights. His current research interests include VLSI testing, design for testability, design verification, dependable computing, and wireless sensor networks. 
    
\end{IEEEbiography}

\begin{IEEEbiography}[{\includegraphics[width=1in,height=1.25in,clip,keepaspectratio]{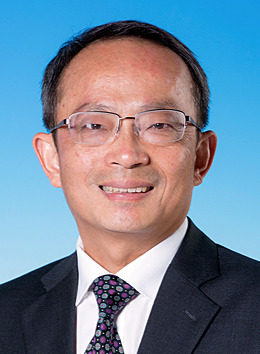}}]{Kwang-Ting Cheng}
    (S’88–M’88–SM’98– F’00) was with the AT$\&$T Bell Laboratories from 1988 to 1993. Before joining HKUST in 2016, he was a Professor of electrical and computer engineering (ECE) with the University of California, Santa Barbara, where he has been since 1993. He is currently the Chair Professor and the Dean of engineering with HKUST. An internationally leading researcher with rich experience in fostering cross-disciplinary research collaboration, his research interests and contributions include VLSI testing and design verification, design automation of electronic and photonic systems, computer vision, and medical image analysis. He had previously served as the Director of the U.S. Department of Defense Multidisciplinary University Research Initiative (MURI) Center for 3-D Hybrid Circuits. He has published more than 450 technical articles, coauthored five books, held 12 U.S. patents, and transferred several of his inventions into successful commercial products.
\end{IEEEbiography}






\end{document}